\documentclass[]{aa}
\usepackage{graphicx}
\usepackage{txfonts}
\usepackage{lscape}
\usepackage{rotating}
\graphicspath{{./figures/}}
\usepackage{natbib}
\usepackage{aalongtable}

\def\etal{et al. }

\def\Ia{SN~Ia~}

\begin{document}
\def\lesim{\stackrel{<}{{}_{\sim}}} \title{Spectroscopy of twelve Type
  Ia supernovae at intermediate redshift} \author{C.
  Balland\inst{1,2,3}, M. Mouchet\inst{2,4}, R. Pain\inst{1}, N. A.
  Walton\inst{5}, R. Amanullah\inst{6}, P. Astier\inst{1}, R. S.
  Ellis\inst{5,7}, S.  Fabbro\inst{8}, A. Goobar\inst{6}, D.
  Hardin\inst{1}, I. M.  Hook\inst{9}, M. J. Irwin\inst{5}, R. M.
  McMahon\inst{5}, J. M.  Mendez\inst{10,11}, P.
  Ruiz-Lapuente\inst{10}, G. Sainton\inst{1}, K.
  Schahmaneche\inst{1}, V. Stanishev\inst{6}}

\institute{LPNHE, CNRS-IN2P3 and Universities of Paris 6 \& 7,
F-75252 Paris Cedex 05, France 
\and APC, 11 Place Marcelin Berthelot, F-75231 Paris Cedex 05, France
\and IAS, University of Paris 11, F-91405 Orsay Cedex,France
\and LUTH, UMR 8102 CNRS, Observatoire de Paris, Section de Meudon, F-92195 Meudon Cedex, France
\and Institute of Astronomy, University of Cambridge, Madingley Road,
Cambridge, CB3 0HA, UK
\and  Department of Physics, Stockholm University, SE-10691 Stockholm, Sweden
\and California Institute of Technology, Pasadena, CA 91125, USA
\and CENTRA-Centro M. de Astrofisica and Department of Physics, IST, Lisbon, Portugal
\and Astrophysics, Denys Wilkinson Building, Keble Road, OX1 3RH , Oxford, UK
\and Department of Astronomy, University of Barcelona, E-08028, Barcelona,
Spain
\and Isaac Newton Group of Telescopes, Apartado 321, E-38700 Santa
Cruz de La Palma, Spain
}

\offprints{christophe.balland@ias.u-psud.fr}

\date{Received; accepted} \titlerunning{Spectroscopy of twelve Type Ia
  supernovae at intermediate redshifts}
\authorrunning{Balland \etal}

\abstract{

We present spectra of twelve Type Ia supernovae obtained in 1999 at
the William Herschel Telescope and the Nordic Optical Telescope
during a search for Type Ia supernovae (SN~Ia) at intermediate
redshift.  The spectra range from $z=0.178$ to $z=0.493$, including
five high signal-to-noise ratio \Ia spectra in the still largely
unexplored range $0.15\leq z \leq 0.3$. Most of the spectra were
obtained before or around restframe B-band maximum light. None of
them shows the peculiar spectral features found in low-redshift
over- or under-luminous SN~Ia. Expansion velocities of
characteristic spectral absorption features such as SiII at $6355$
\AA, SII at $5640$ \AA ~and CaII at $3945$ \AA ~are found consistent
with their low-$z$ \Ia counterparts.
  
   }

\maketitle

\section{Introduction}

Type Ia supernovae (SN~Ia) provide us with a powerful tool for
constraining cosmology through the magnitude-redshift Hubble diagram.
In the last decade, systematic searches for \Ia have been performed in
view of increasing the statistics at all redshifts, while at the same
time trying to improve our understanding of these events.  Programs
launched in the early nineties were targeting high-redshift
($0.4\lesim z\lesim0.8$) supernovae and have lead to the discovery of
the acceleration of the expansion of the
universe~\citep{Riess98b,Perlmutter99}.  More recently, even more
distant \Ia (up to redshift $\sim1.7$) have been discovered and
followed-up with HST/ACS allowing one to ``observe'' the turn-around
epoch of the acceleration history of the universe \citep{Riess04}.
The nearby supernovae ($z\lesim 0.1$) have also received great
attention, as they are necessary to anchor the Hubble diagram at low
redshift and because they offer the possibility to study systematics
since they are easier to observe than their high redshift
counterparts.

Although the number of high-redshift \Ia discovered and followed-up
greatly increased along the years and is now rapidly approaching a
thousand\footnote{current high-statistics programs such as ESSENCE
  (www.ctio.noao.edu/$\sim $wsen) or SNLS (cfht.hawaii.edu/SNLS) now
  target number statistics of well measured \Ia well above the several
  hundreds in large multi-year projects}, the gap observed in the
Hubble diagram at intermediate redshift range has endured.  Indeed,
optimizing supernova searches in that redshift range requires securing
large fractions of observing time on an intermediate size telescope
($\sim2m$) with a large field of view ($\ge 0.25^{o2}$) in order to
ensure {\it efficient} detection of a large enough number of
supernovae for the duration of the survey.  Intermediate redshift
supernova programs are thus subject to the availability of such
telescopes and instruments.  Nevertheless, observing supernovae at
intermediate redshift is crucial not only to "bridge the gap" between
the low- and high-redshift samples to increase the precision in
measuring the cosmological parameters, but also to obtain precise
restframe U-band photometry and spectroscopy of SN~Ia, otherwise
difficult to obtain on nearby objects because of the low detector
efficiency and strong variability of the atmosphere in the U-band.
  
Of utmost importance is to check that the properties of \Ia do not
significantly evolve with redshift.  Among potential systematics is
the evolution of SNe Ia spectral characteristics with $z$.
Differences in the SNe Ia environment and parent population are likely
to translate into differences in the presence and evolution of
spectral features.  Indeed, as our knowledge of observable properties
of \Ia improves, it clearly appears that diversity rather than
uniformity is common among type Ia supernovae
\citep{Branch87,Branch88,Hatano00,Benetti04,Benetti05}.  It is crucial
to quantify this diversity in order to estimate to what extent it
affects the validity and accuracy of the cosmic parameter
determination from the Hubble diagram.  In this respect, an enlarged
sample of precisely measured intermediate redshift \Ia spectra is
needed to improve our understanding of their behavior.

It is thus important to discover and follow-up both photometrically
and with spectroscopy a substantial number of \Ia in the intermediate
($0.1<z<0.4$) redshift range. For this reason we launched, during the
year of 1999, a campaign to search and follow-up \Ia in that redshift
range. A second campaign was pursued during the year 2002.  In this
paper, we present the spectra taken at the William Herschel Telescope
(WHT) and at the Nordic Optical telescope (NOT) of the twelve
supernovae discovered in the spring and fall 1999, at the Isaac Newton
Telescope in La Palma observatory. The photometric follow-up of five
of these supernovae and their inclusion in the Hubble diagram will be
presented in a subsequent paper. In Sect. 2, we briefly describe the
search strategy and discovery technique used.  In Sect. 3, the
spectroscopic observations of the supernova candidates, including the
reduction procedure, are presented. In Sect. 4 and 5, we focus on the
identification and analysis of our set of \Ia spectra. Discussion of
results and comparison to other works are done in Sect. 6.  Concluding
remarks are given in Sect. 7.

\section{Search for \Ia at intermediate redshift}

As the present paper focuses on the spectral analysis of the \Ia
sample, we only briefly describe here the strategy and technique used
to discover the supernovae.

\subsection{Overall strategy}
 
The search for supernovae was performed on images taken using the Wide
Field Camera (WFC), a four 2k$\times$4k thinned EEV CCD mosaic with a
$\sim 0.27$ square degree FOV mounted on the Isaac Newton telescope. Most of
the observations were done as part of the Isaac Newton Group's Wide
Field Survey (WFS), a public access survey of multicolor data obtained
over a 5 year period through the INT photometric filters set (u'
through z') over more than 200 square degrees with a typical depth of
$r'\approx 24$ and $g'\approx 25$ \citep{Walton99, McMahon01}. WFS
observations were scheduled as to permit repeated observations of the
same area of sky.  Detection of the SN candidates was done by repeated
imaging, in the Sloan-Gunn g'1996 filter, of the South Galactic Cap
(SGC) fields over an area of $\sim 31$ square degrees.  SGC fields at
zero declination were chosen so that follow-up from both northern and
southern hemispheres could be done.  After a test period in April 1999
whose purpose was to validate our strategy and detection technique,
600 s images were taken around the new moon in August 1999 (new moon
on August 11th) and then used as the input baseline {\em reference}
images for the subsequent supernova search. As the restframe rise
time for \Ia is about 18 days \citep{Aldering00,Riess99b}, two new
images of 240~s each of the same fields in the same g' filter were
taken roughly a month later (around the September new moon) in order
to discover the supernovae around their maximum light
\citep{Perlmutter97}.  The schedule of this new set of observations
was optimized to ensure the best matching in seeing between the
reference and new images. The image quality of the data was generally
good, with a median seeing of $\approx 1$ arcsec.

\subsection{Discovery technique}

The search procedure generates potential \Ia candidates, as well as
other objects, such as AGN, other SNe types, asteroids, etc. The new
image (the discovery image) consisted in two exposures taken one hour
apart, so as to reject cosmic ray hits and asteroids. After
pre-processing , the discovery images were co-added and three
subtracted frames were built, by subtracting, from the reference
image, one or the other discovery image, or their sum. These
subtracted frames were then searched for point-like sources. The
supernova candidates were selected by requiring that they were
detected on the 3 subtracted frames, at the same place within 2 pixels
(0.66 arcsec) in order to reject cosmic rays and asteroids, and that
their flux within an aperture of 1 full width at half maximum (FWHM)
in radius is at least larger than 15 \% of the flux measured in a
similar aperture on the reference frame (so as to eliminate
subtraction residuals). This requirement also rejects slow varying
objects, as plateau type II supernovae.  Finally, the supernova
candidates were visually checked by two independent scanners. These
candidates were re-observed the following night, and kept only if
re-detected with a similar (or higher) flux.  This last procedure
tends to eliminate any slow-moving object and supernovae on the
decline. The quality of the candidate as a potential supernova was
estimated from its flux and shape parameters and prioritized as very
likely, probable or possible supernova. A total of fifteen candidates
was obtained (the 'September candidates' \citet{Astiera}).  Spectra
were obtained within a few days after detection during pre-scheduled
spectroscopic follow-up time at WHT.  For two candidates,
complementary spectra were obtained at the Nordic Optical Telescope
(NOT) about three weeks after discovery.  Ten out of the fifteen
spectroed candidates have been confirmed as \Ia. To this set, we add
two more supernovae (the 'April candidates' \citet{Hardina,Hardinb})
obtained in a similar way during the test run in April 1999 (see
below).

\section{Spectral observations and reduction}

\subsection{Observations}

Spectra of all the supernova candidates were obtained with the ISIS
spectrograph at the 4.2-m William Herschel telescope (WHT) on April 19
and 20 (SN~1999cj and SN~1999ck discovered during the test run) and on
September 12th, 13th and 14th 1999 for the fifteen September
candidates.

For April 1999 candidates, both the red and blue channels of ISIS were
used along with the ISIS 6100 dichroic filter.  For the blue part, we
had a blue sensitive thinned EEV CCD of 4096$\times 2048$ pixels
combined with the grating R158B (useful spectral range from 4000 to
6200 \AA, dispersion 1.62 \AA$\ $ per pixel). For the red part, the
detector used was a $1024 \times 1024$ pixel TEK CCD device operating
in its standard mode and combined with the R158R grating covering a
total useful spectral range of $\sim$ 6000 to 8000\AA$\ $with a
wavelength bin per pixel of 2.90 \AA.  With this device, fringing is
negligible below 7500 \AA.

Three exposures of 1800s each were obtained for the two April
candidates, both in the blue and red parts, with a long slit of 1
arcsec width.  Copper-neon arcs were used for wavelength calibration.
Two standards were observed and used for flux calibration
(BD+33$^\circ$2642 and BD+28$^\circ$4211, see below).

During the September campaign, only the red channel of the
spectrograph was used in order to avoid some sensitivity loss due to
the dichroic filter. Even though the two April spectra were of
acceptable quality, removing the dichro\"{\i}c filter allowed us to
optimize the efficiency in the red channel, where the principal
features of \Ia are present for our redshift range.  The red arm had
the same detector as in April.  Three long-slit spectra of 1800\,s\ 
each were obtained for each candidate, except for SN~1999dx, SN~1999dv
and SN 1999gx for which only two 1800\,s exposures were taken. A slit
width of $\sim$1.5 arcsec was used.  Parallactic angle was generally
used to minimize loss by differential refraction. In some cases, the
slit was put at a specific orientation in an attempt to take a
spectrum of the host galaxy at the same time as the supernova
spectrum.  Three spectrophotometric standards were observed each night
(Feige~25, BD+28$^{\circ}$4211 and Feige~110), with both $\sim$1.5
arcsec and 8 arcsec widths.  Copper-neon arcs were obtained at the
beginning and end of each night for the purpose of wavelength
calibration.

Weather conditions were good for the three September nights with a
seeing of about 1 arcsec or less, except for the first half of the
second night (seeing about 3 arcsec). Five candidates were observed
during each night. Two additional exposures of one faint candidate
(SN~1999dz first observed on September 13) were obtained during the
night of September 14th.

Additional observations of SN~1999du and of SN~1999dv were obtained on
October 2nd, 1999 at the 2.5m Nordic Optical Telescope (NOT) in La
Palma.  Long slit spectra have been acquired with the ALFOSC
spectrometer, using a 300 lines/mm grism (number 4) and a CCD Loral
(2048x2048 pixels) detector of 15 $\mu m$ pixel size.  The wavelength
coverage ranges from 4200 to 8900 \AA. However, due to the presence of
the second order spectrum with this grism, the useful range is reduced
to approximately 4200 to 6000\AA.  A slit width of 2.5 arcsec and of 1
arcsec, yielding a resolution of 35-40 \AA{} and of 14 \AA$\ ${} have
been used for SN~1999du and SN~1999dv respectively.  Five spectra of
SN~1999du and six spectra of SN~1999dv were obtained with an exposure
time of 1800\,s.  The last spectrum of SN~1999du is very weak and not
considered in the following.  BD+28$^{\circ}$4211, observed on October
3, has been used for flux calibration purposes and helium lamp spectra
were taken for wavelength calibration.  The seeing during the NOT
observations was always better than 1 arcsec.

A summary of the spectroscopic observations are given in Table
\ref{tab1}.

\subsection{Reduction}

A preliminary assessment of the spectroscopic data was performed
interactively at the telescope to allow for confirmation of the
candidate, so as to select the supernova sample for photometric
follow-up as soon as possible after discovery.

The spectroscopic reduction was redone using the ESO-MIDAS data
reduction software version 02SEP.  An average bias and normalized flat
field frames, obtained from an internal tungsten lamp, are produced
for each night using a median filter in order to remove cosmic ray
hits.  A two dimensional wavelength calibration is done, using
copper-neon arcs, by computing a dispersion relation for each row of
pixels along the dispersion direction (horizontal with our set-up)
over the entire two dimensional frame.  This corrects for some
distortion along the wavelength direction and minimizes subtraction
residuals after sky removal. About 40 lines are identified and used,
covering the total spectral range in a regular way. This leads to a
precision better than 0.05\% over the whole spectrum. Finally, we
check on a few prominent sky emission lines (NaI at 5893\AA$\ $ and
[OI] at 5577, 6300 and 6363 \AA) that the wavelength accuracy is
consistent with this value, corresponding to less than 1 \AA$\ $
difference between the calibrated sky line and its theoretical
wavelength.  Extraction of the spectrum is then performed from the 2D
calibrated frame using a procedure based on \citet{Horne86} weighting
algorithm optimizing sky subtraction and flux extraction. An error
spectrum, essentially dominated by the sky signal, is generated.

After correcting for atmospheric extinction using the extinction curve
of La Palma observatory \citep{King85}, absolute flux calibration is
done. A response curve is computed by comparing our observed
spectrophotometric standard stars with their tabulated absolute flux
(in practice, we used Feige 110 to calibrate all the September
spectra, and BD+28$^\circ$4211 and BD+33$^\circ$2642 for,
respectively, the blue and red part of the April spectra).  This
response curve is then applied to our spectra. Finally, we remove ``by
hand'' sky residuals resulting from bad sky subtraction or remaining
cosmics.  We do not attempt to remove atmospheric absorption lines
around 6900 \AA{}.

For the two April spectra, the reduction and extraction was performed
on the blue and red frames in a separate way. After flux calibration,
the two parts of the spectrum were combined to give the final full
spectrum, with a useful wavelength coverage from about 4000 to 7500
\AA.  The two parts overlap nicely over $\sim 200$ \AA$\ $ around 6100
\AA$\ $, showing that our flux calibration or extraction procedure is
not significantly in error.

As for the WHT data, standard reduction for NOT data was performed in
the ESO-MIDAS environment (cosmic removal, bias subtraction,
flat-field correction, 2D-wavelength calibration, and flux
calibration).  Again, the wavelength calibration was checked against
sky lines. They were found within 3-4 \AA{} from their tabulated
wavelengths.  Extraction was done following the same procedure as for
WHT spectra.

\section{Identification}

\subsection{Redshift determination}

Redshift determination is based on the identification of
characteristic galactic emission and absorption lines which are
narrower than any supernova feature. We use mainly [OII] at 3727\AA,
CaII H\&K at 3945 \AA, [OIII] doublet at 5000\AA\ or $H\beta$, or
several of these lines together if present.  We do a gaussian fit of
the line profile.  The relative spectral resolution $\delta
\lambda_{obs}/\lambda_{obs}$ is typically less than 0.05 \% over the
useful spectral range (see Sect. 3.2) which translates into an
absolute error on the redshift $\delta z = \delta
\lambda_{obs}/\lambda_{obs}\times (1+z)$ less than 0.001 at $z=0.5$
and less than 0.0005 at $z \sim 0.1$. In the following, redshifts are
thus given with three significant figures.

\subsection{SN template fitting}

Identification of the candidates is done using a software tool
developed by one of us \citep{Sainton04a,Sainton04b}.  The software,
called ${\cal SN}$-fit, is based on a $\chi^2$ minimization fitting of
the reduced spectrum to a set of known spectral templates. As
contamination from the host galaxy often occurs, a model spectrum
${\cal M}$ consisting of the weighted sum of a fraction of a supernova
${\cal S}$ and a fraction of a galaxy template ${\cal G}$ (or of the
observed galaxy host when available) is built in the restframe:
\begin{equation}
\label{eq1}
{\cal M}(\lambda_{rest},z,\alpha,\beta)=\alpha{\cal S}(\lambda_{rest}(1+z))
+\beta{\cal G}(\lambda_{rest}(1+z))
\end{equation}
All template spectra are initially in the restframe and are
appropriately redshifted. The redshift determined from galaxy lines
(see previous subsection) is used to constrain the input redshift
interval to search for the solution, although it is not mandatory as
the input redshift interval can be left arbitrarily large ($z$ is then
considered as a free parameter).  The $\chi^2$ fitting is done against
the set of parameters ($z$,$\alpha$,$\beta$) and is robustified
against undesirable points such as regions where atmospheric
absorption takes place or where sky subtraction residuals are present.

The fit is done for every template couple (supernova, galaxy)
available in our database and results are classified in increasing
order of reduced $\chi^2$. In order to assess the significance of the
best-fit solution, we systematically examine the immediately following
solutions to check for possible degeneracies in the models.  It is
clear that the result will greatly depend on the content of the
database and special care has been put to build it. It contains about
150 spectra (mostly supernovae plus a dozen galaxy templates) taken
from various sources.  The set of supernova templates contains local
'normal' \Ia as well as several peculiar Ia and Ib/Ic/II supernovae.
These spectra have been put into the restframe. Galaxy templates are
taken from \citet{Kinney96}.  Morphological types include bulge, E,
S0, Sa, Sb, Sc, and starbursts (Stb) with various amount of reddening
\citep{Calzetti94,Kinney96}. We refer the reader to Appendix A for
more details.

One concern with the database is the uniformity of the sampling of the
supernova phase (the date of the SN spectrum template with respect to
maximum light).  If the sampling is too scarce, this limits the
resolution on the phase of the observed spectrum we get from the fit
(we call this phase the ``spectral phase''). We find a typical $\pm 3$
day difference on the spectral phase. Higher differences between two
solutions similar in terms of $\chi^2$ can happen when the SN signal
is weak with respect to the host galaxy or when the spectrum of the SN
is acquired well after maximum light, as spectral features evolve less
rapidly a few weeks after maximum (phase degenerate solutions). Since the
detection procedure is optimized to pick up SN around maximum, this
latter case happens only rarely.  Another concern is related to the
fact that we do not try to make any accurate determination of the
reddening to the SN.  Indeed many effects combine to account for the
reddening that are difficult to separate: differential slit losses
when not at the parallactic angle, errors in flux calibration and
template spectra uncorrected for reddening.  Modeling 'reddening' by
adding a term in equation (1), as in \citet{Howell02}, is thus subject
to caution as interpretation of such a term is difficult and we choose
not to do so.  This means that some degeneracy might exist between the
best-fit phase and a possibly 'reddened' SN or template.  This might
increase the uncertainty in the phase determination. 

To quantify this effect, we have artificially reddened the template
spectrum of SN~1994D at -2 days \citep{Patat96}, using the reddening
law of Howarth \citep{Howarth83} for E(B-V)=0.2, 0.4, 0.6 and 0.8. We
have then fitted this spectrum, using ${\cal SN}$-fit, with a set of
spectral templates introduced by \citet{Nugent02} and modified by
\citet{Nobili03}. These templates are constructed from a broad sample
of \Ia and they are calibrated so as to reproduce the magnitude
evolution along the supernova light-curve in the BVRI passbands.
Spectrally, they offer, contrary to local templates of individual \Ia,
a phase resolution of 1 day over a large phase range.  For our
purpose, we only consider templates between -10 and + 10 days. For
each value of E(B-V), we compute the phase from the first best-fit
solutions by weighting the phase $\phi_i$ of each solution with the
corresponding $\chi^2$ probability $p_i$. For the unreddened spectrum
of SN~1994D, as well as for E(B-V)=0.2, the phase obtained is $\approx
-3.5$ days.  The original -2 days phase is thus recovered within less
than 2 days.  For E(B-V)=0.4, 0.6 and 0.8, the difference between the
original phase and the determination from ${\cal SN}$-fit increases:
the obtained phase is $\approx +3.5, +5.5$ and $+9$ respectively.  As
the original spectrum is progressively reddened, later-phase (i.e.
redder) spectra are selected by the identification procedure. Note
that the visual aspect of the fit is considerably degraded for E(B-V)
values higher than 0.2. We conclude that for moderate reddening, the
phase determination is reliable within 2 to 3 days.  For higher
values, no satisfying fit is obtained. In these cases, visual
examination of the spectrum reveals the reddened nature of the
spectrum.
  
An other concern is the possible misidentification of a \Ia spectrum
around maximum as a type Ib/Ic or II supernova.  The P-Cygni features
of a type II spectrum might fit some portion of the red part of a \Ia
spectrum before or close to maximum.  More likely, a type Ic spectrum
might be misidentified as a type Ia.  In particular, this can happen
when the fit is performed on a too small portion of the spectrum to be
identified.  In any case, it is important to have a sufficient number
of type Ib/Ic/II spectral templates in the database. At the time of
this analysis, we dispose of 64 spectra of type Ib/Ic/II objects (see
Appendix A).
  
In the present analysis, the redshift is always determined from galaxy
lines and it can be safely constrained in ${\cal SN}$-fit.  Templates
selected in the spectral database typically cover a large wavelength
range but as they are redshifted, the overlap between the model
spectrum (redshifted SN + Galaxy) and the observed spectrum might be
small.  This can lead to an unreliable identification as confusion
with other types may occur. It is thus important to check {\it a
  posteriori} that the best-fit solution spans a large enough
wavelength range.  In case it is too small, inspection of adjacent
solutions is necessary to assess the validity of the best-fit
solution.
  
Obviously, the reliability of the identification depends on the
underlying model, which is directly connected to the number, quality
and variety of templates available in the database. A good phase
sampling is necessary but a sufficient number of spectra of different
\Ia supernovae at a given phase is also required in order to reflect
the intrinsic differences observed in \Ia spectra at low redshift.
Note that the higher the quality of the observed spectrum is, the more
significant the discrepancies between the model and the observed
spectrum are, in terms of $\chi^2$ value.  A high best-fit $\chi^2$
value for a given spectrum either happens when the observed spectrum
does not correspond to a supernova (and thus no satisfying model can
be constructed from templates available in the database), or when the
signal-to-noise of the observed spectrum is high. In any case,
$\chi^2$ changes between solutions for a {\it given} supernova are
meaningful but direct comparison of $\chi^2$ best-fit values for {\it
  different} supernovae should be done only in the case of comparable
signal-to-noise.

\subsection{Host galaxy subtraction}

Host galaxy subtraction is clearly an important step in the supernova
spectrum identification. In our sample, some of the supernova spectra
are indeed deeply contaminated by the emission of their host. This is
the case for SN~1999dt, SN~1999du (WHT spectrum), SN~1999dw, SN~1999dx,
 SN~1999dz or SN~1999gx.  Identification of the host type is done on the full
(i.e. SN+host) extracted spectrum by direct examination of several
spectral features allowing to discriminate, at least broadly, between
types. We adopt a division into three main morphological classes,
similarly to \citet{Sullivan03}: type 0 corresponds to spheroids
(E/S0/bulge), type 1 to early-type spirals (Sa/Sb) and type 2 to
late-type spirals and starbursts (Sc/Stb). (We do not have any
irregular galaxy template in our database.) Spectral features used for
the identification (as well as for redshift determination; see \S 4.1)
include CaII H\&K absorption lines at $3934$ and $3968$ \AA, the 4000
\AA$\ $ break ($B4000$), Hydrogen Balmer lines (mostly $H\beta$ and
$H\gamma$ given the redshift range of the present spectra), oxygen
forbidden lines [OII] and [OIII].  We also consider a feature around
$3850$ \AA: the presence in early-type spectra of a large trough,
absent in later-type spectra, due to CN molecular absorption blueward
of $3850$ \AA\, in cool stars (we label it T3850). This feature
correlates with the strength of Mgb absorption at $\sim 5150$ \AA$\ $
(restframe), an other useful metallicity indicator. Using all these
features allows classification of all host galaxies into one of the
three main categories defined above.

Subtraction of the host from the supernova spectrum is performed as
part of the fitting procedure by ${\cal SN}$-fit. In two cases
(SN~1999dr and the NOT spectrum of SN 1999du), it is possible to use
the observed host spectrum to create the model to be fitted.
Subtraction of galaxy lines is then optimal. In all other cases,
however, it is not possible to extract separately the host. We then
use a host galaxy template of high signal-to-noise ratio appropriately
redshifted (see \S 4.2).  This offers two advantages over a
subtraction based on using the real host spectrum: first, as the
template used has a high S/N ratio, the resulting supernova S/N is
essentially not degraded by the subtraction; second, as the extraction
procedure relies on a $\chi^2$ minimization, the best-fit solution
gives in itself a hint of the host type (the 'best-fit host type')
that can be used as an {\em a posteriori} identification check by
comparing to the identification based on galaxy features.  Moreover,
the ratio of the galaxy to the total supernova+galaxy signal is
computed by ${\cal SN}$-fit as a byproduct of the fitting procedure.
This ratio represents the contribution of the galaxy in the best-fit
model in the range of overlap with the observed spectrum. This allows
us to quantify the degree of contamination by the host. Although
limited by the imperfect sampling of galactic type templates in our
database, it nevertheless gives a valuable indication on the host type
and degree of contamination of the supernova.

\section{Spectroscopic results}

\subsection{Notes on individual \Ia}

As an output of the fitting procedure, a host-subtracted supernova
spectrum is thus obtained along with the best-fit supernova template
and parameters. Table \ref{tab2} summarizes these results of the
fitting procedure for each supernova. Columns 3 and 4 give the
best-fit supernova template along with the percentage of galaxy
contamination and galaxy type. Columns 5 and 6 give the best-fit
redshift $z_f$ obtained from the fitting procedure and the host
redshift $z_{h}$ respectively. $z_f$ is obtained by redoing a fit of
the supernova spectrum obtained after a first fit, cleaned from any
residual galactic line.  Column 7 gives the phase associated with the
best-fit supernova+galaxy spectrum. As mentioned above, the phase has
to be considered with a typical $\pm 3$ day uncertainty.  Finally,
the minimum reduced $\chi^2$ and number of degrees of freedom are
given in columns 8 and 9.

Examining, for each supernova, the solution immediately following the
best-fit solution in the list of increasing reduced $\chi^2$, we
generally find that the phase is rather stable and within $\pm 3$ days
of the best-fit solution. For three supernovae (SN~1999cj, SN~1999dr
and SN~1999gx), the phase difference is larger, up to $\pm 5$ days.
Note that SN~1999dr is a supernova well after maximum light (its
spectral features evolve less rapidly than around maximum).  The same
argument may apply to SN~1999cj (+9 days), although the effect is not
found for the +9 days SN~1999dv NOT spectrum.  The signal-to-noise of
this latter spectrum is however higher than for SN~1999cj and phase
determination should be more accurate. Finally, SN~1999gx is a distant
SN and is dominated by the host galaxy.

\begin{figure*}
\resizebox{14cm}{!}{\includegraphics{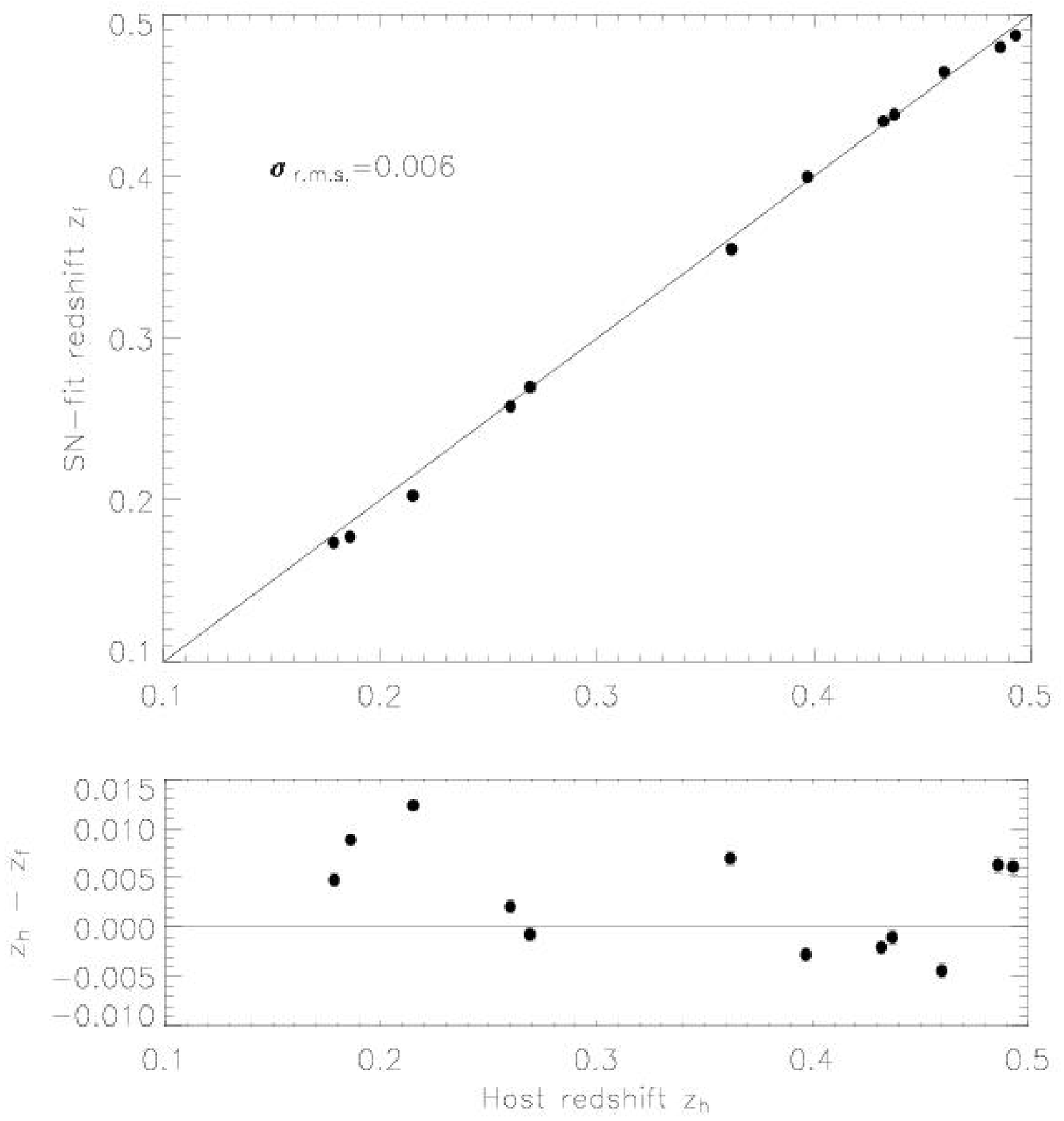}}
\caption{Redshift $z_f$ from ${\cal SN}$-fit as a function of the
  redshift $z_h$ of the host galaxy (top panel), and residuals
  (bottom panel). Error bars shown are for the host redshift only.}
\vfill
\label{fig1}
\end{figure*}

The best-fit solutions are obtained in all cases for 'normal' \Ia
templates, as opposed to peculiar \Ia and other types (Ib/c and II).
Figure \ref{fig1} (upper panel) shows the best-fit redshift $z_{f}$
(column 5 of Table \ref{tab2}) as a function of the redshift obtained
from line identifications (the host redshift $z_h$, column 6 of Table
\ref{tab2}).  Residuals are shown on the lower panel of Fig.
\ref{fig1}.  The two quantities agree within a few percent. Error bars
on the host redshift are shown (they are smaller than the filled
circle symbols and do not appear on the upper panel). When galactic
features are weak, more difference is observed (see, e.g., SN~1999dy)
but no clear systematic effect has been identified (see lower panel of
Fig. \ref{fig1}).  We find a r.m.s.  dispersion of $\sigma_{r.m.s}
\approx 0.006$ for the whole sample.  This value can be considered as
a typical error on the redshift determination by ${\cal SN}$-fit. The
average redshift of the sample is $<z>=0.348$.

\begin{figure*}[t]
\resizebox{14cm}{!}{\includegraphics{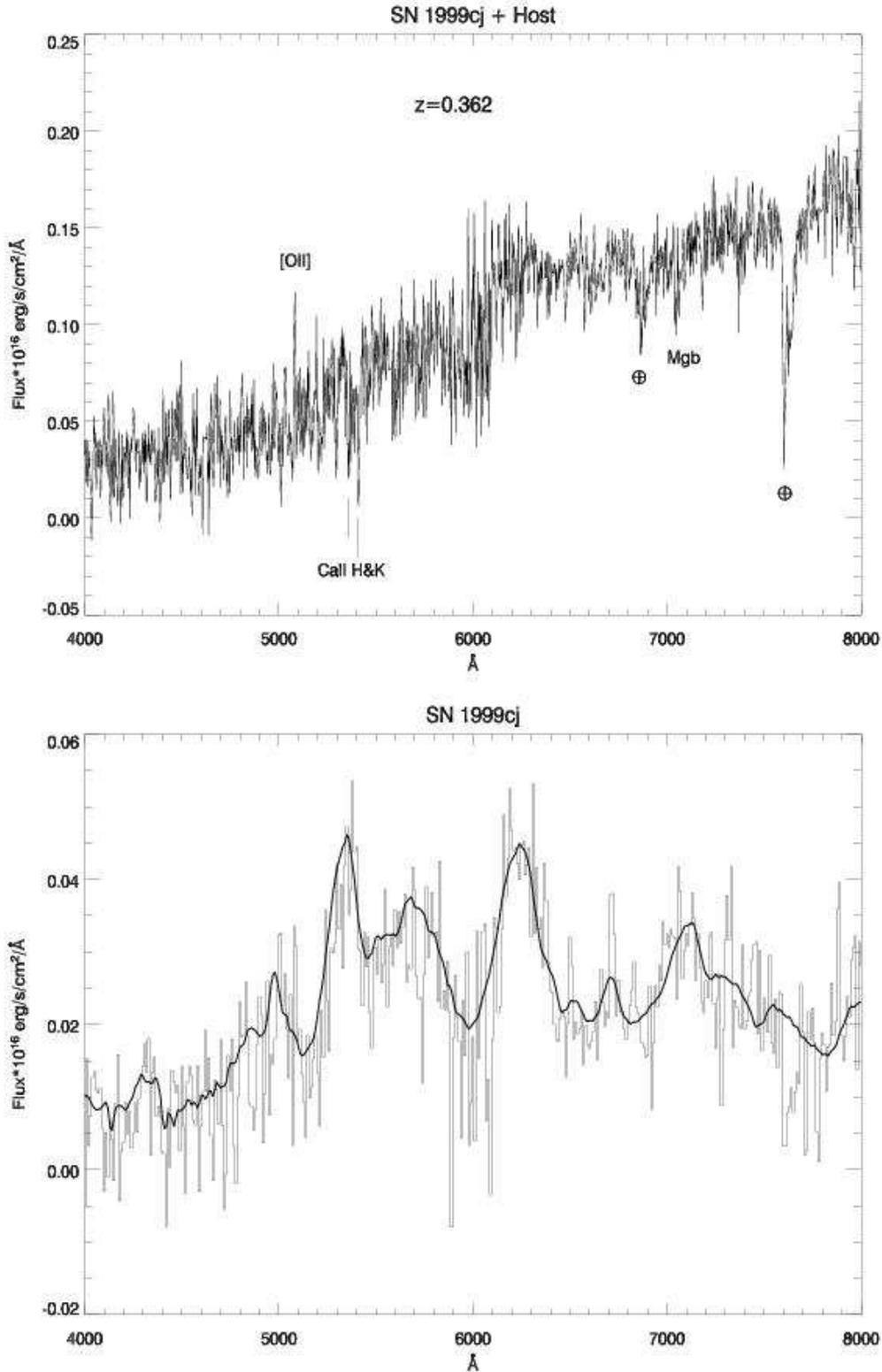} }
\caption{Top: SN~1999cj+host in the observer frame. Bottom: 
  Galaxy-subtracted SN~1999cj, rebinned for visual convenience, 
  with the best-fit template (SN~1992A +9
  days) overlapped. The spectrum is shown in the observer frame and is
  not corrected for atmospheric absorptions or galactic line
  subtraction residuals. Note the restframe UV part accessible due to
  the use of ISIS blue arm for this supernova.}
\label{fig2}
\end{figure*}

\begin{figure*}
 \resizebox{14cm}{!}{\includegraphics{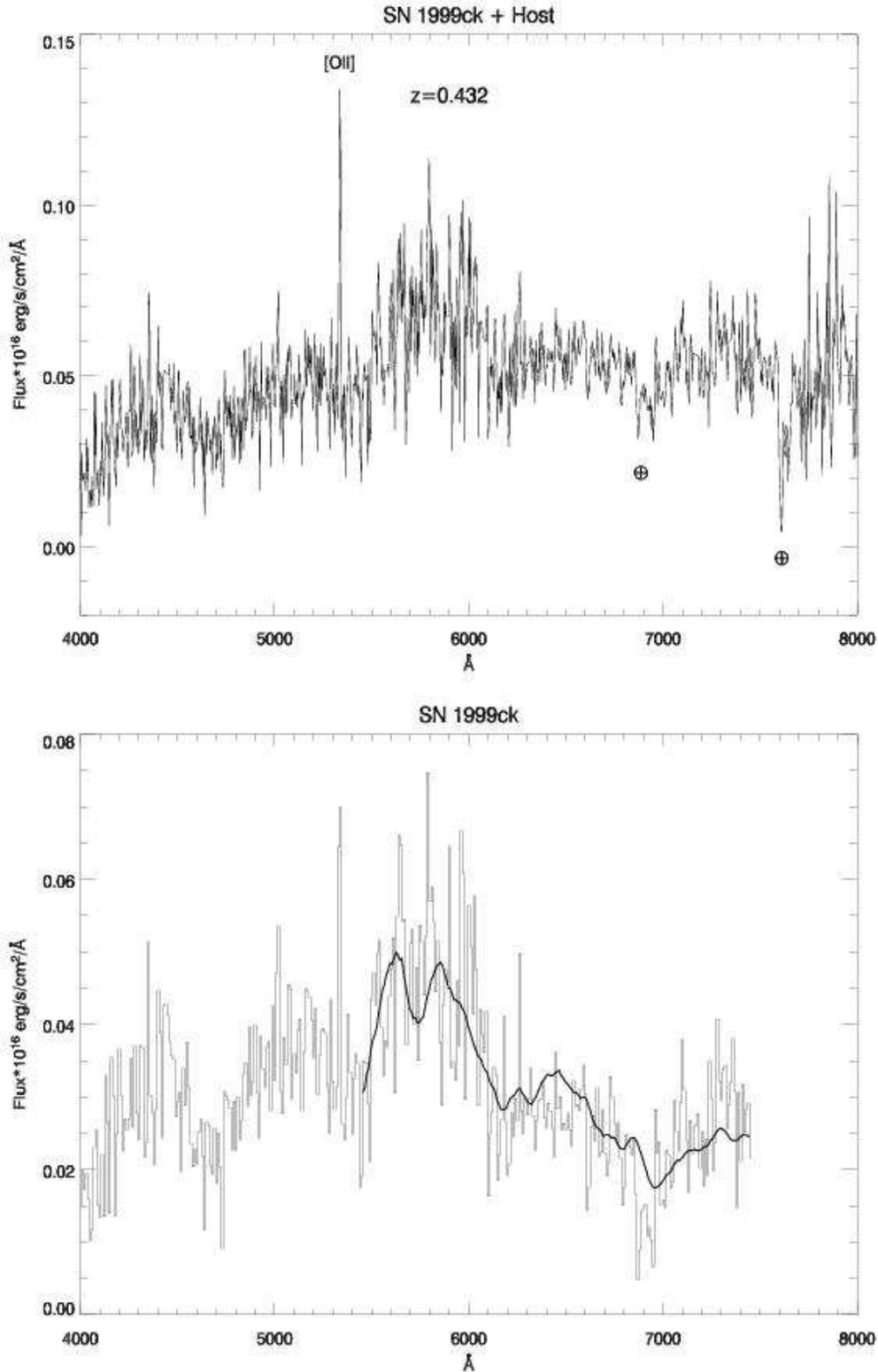} }
\caption{Top:  WHT spectrum of SN~1999cj+host in 
  the observer frame. Bottom: Rebinned galaxy-subtracted spectrum of
  SN~1999ck with the best-fit template (SN~1994D -9 days) overlapped.
  The spectrum is shown in the observer frame and is not corrected for
  atmospheric absorptions or galactic line subtraction residuals. As
  for SN~1999cj, the restframe UV part is visible due to the use of
  ISIS blue arm.}
\label{fig3}
\end{figure*}

\begin{figure*}
 \resizebox{14cm}{!}{\includegraphics{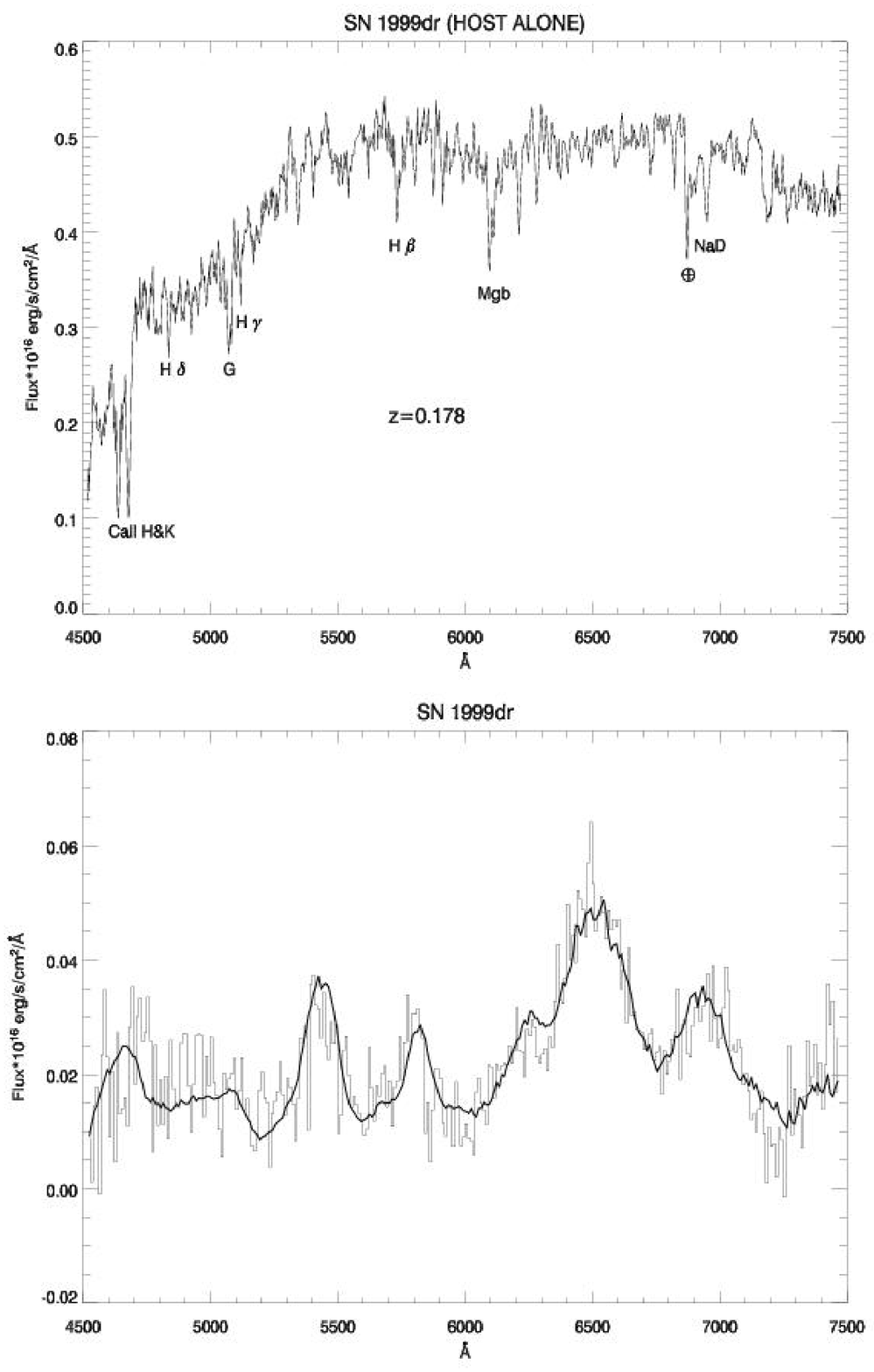} }
\caption{Top:  WHT spectrum of SN~1999dr host galaxy 
  alone. Bottom: Rebinned galaxy-subtracted spectrum SN~1999dr with
  the best-fit template (SN~1994D + 24 days) overlapped. The spectrum
  is shown in the observer frame and is not corrected for atmospheric
  absorptions or galactic line subtraction residuals.}
\label{fig4}
\end{figure*}

\begin{figure*}
\resizebox{14cm}{!}{\includegraphics{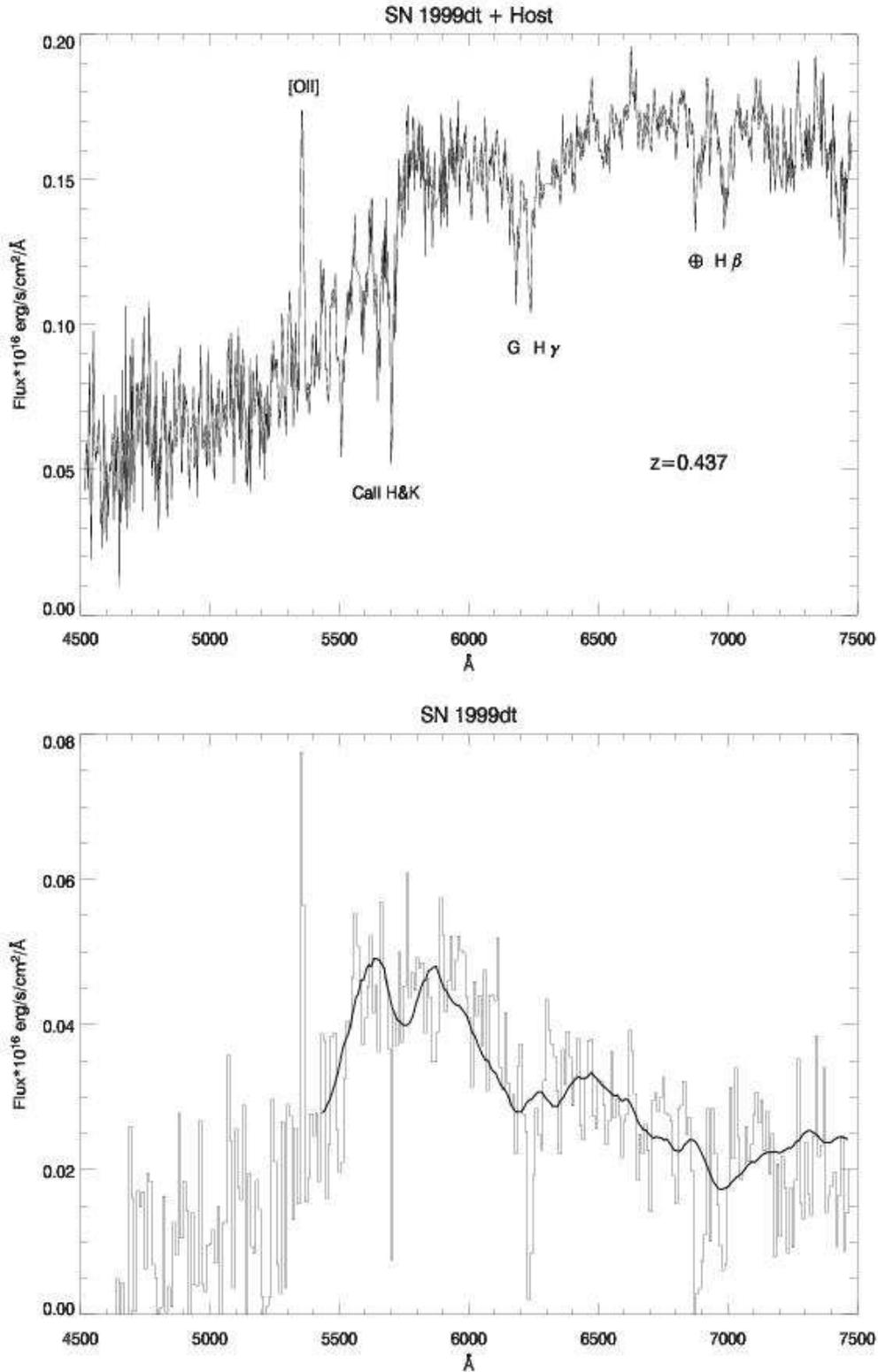} }
\caption{Top: WHT spectrum of SN~1999dt+host. 
  Bottom: Rebinned galaxy-subtracted spectrum of SN~1999dt with the
  best-fit template (SN~1994D - 9 days) overlapped. The spectrum is
  shown in the observer frame and is not corrected for atmospheric
  absorptions or galactic line subtraction residuals. Note the poor
  [OII], $H_\beta$ and $H_\gamma$ line subtraction and the presence of
  strong atmospheric absorptions.}
\label{fig5}
\end{figure*}

\begin{figure*}
  \resizebox{14cm}{!}{\includegraphics{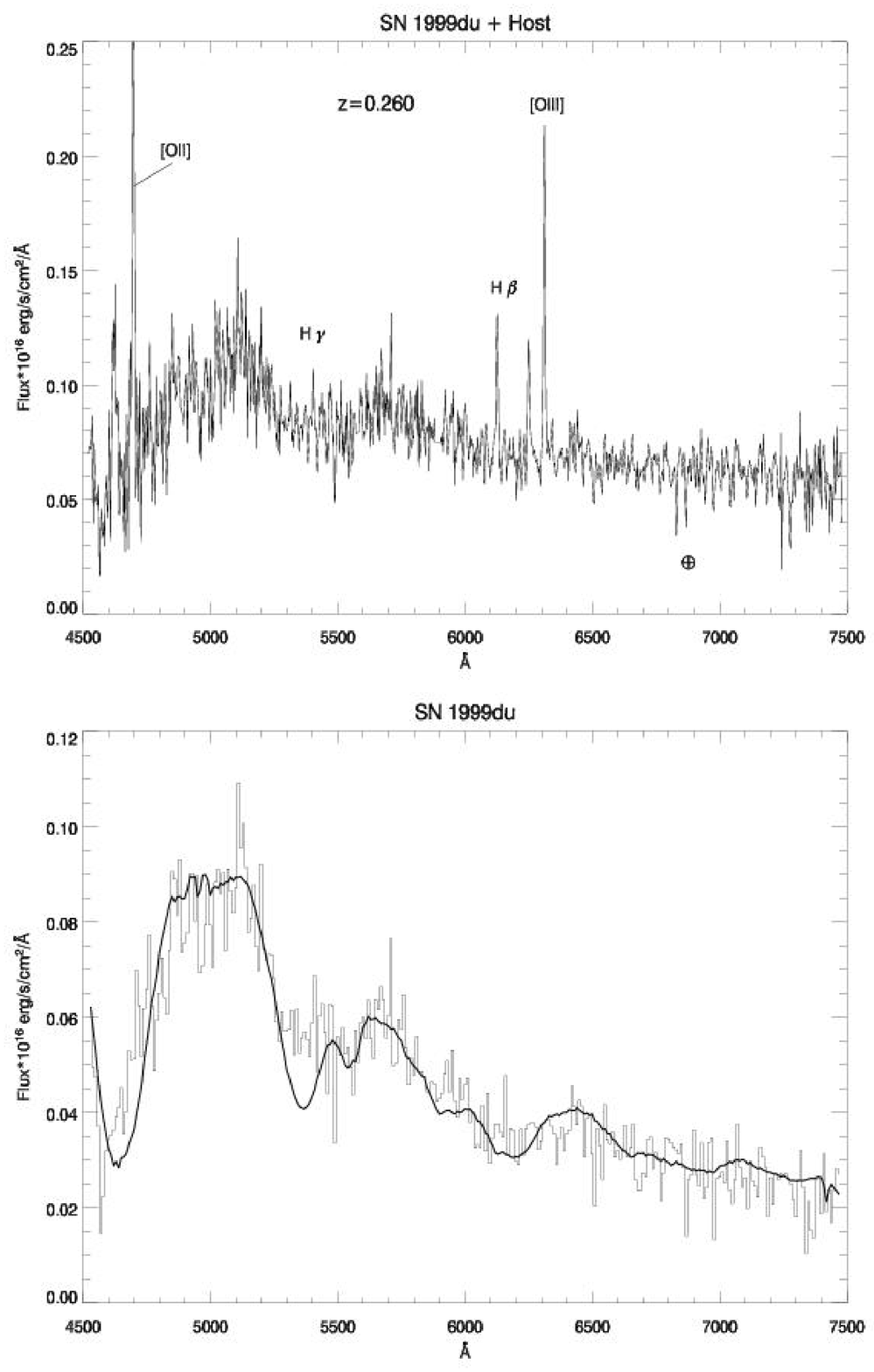} }
\caption{Top: WHT spectrum of SN~1999du+host. 
  Bottom: Rebinned galaxy subtracted spectrum of SN~1999du with the
  best-fit template (SN~1999ee -9 days) overlapped. The spectrum is
  shown in the observer frame and is not corrected for atmospheric
  absorptions or galactic line subtraction residuals.}
\label{fig6}
\end{figure*}

\begin{figure*}
 \resizebox{14cm}{!}{\includegraphics{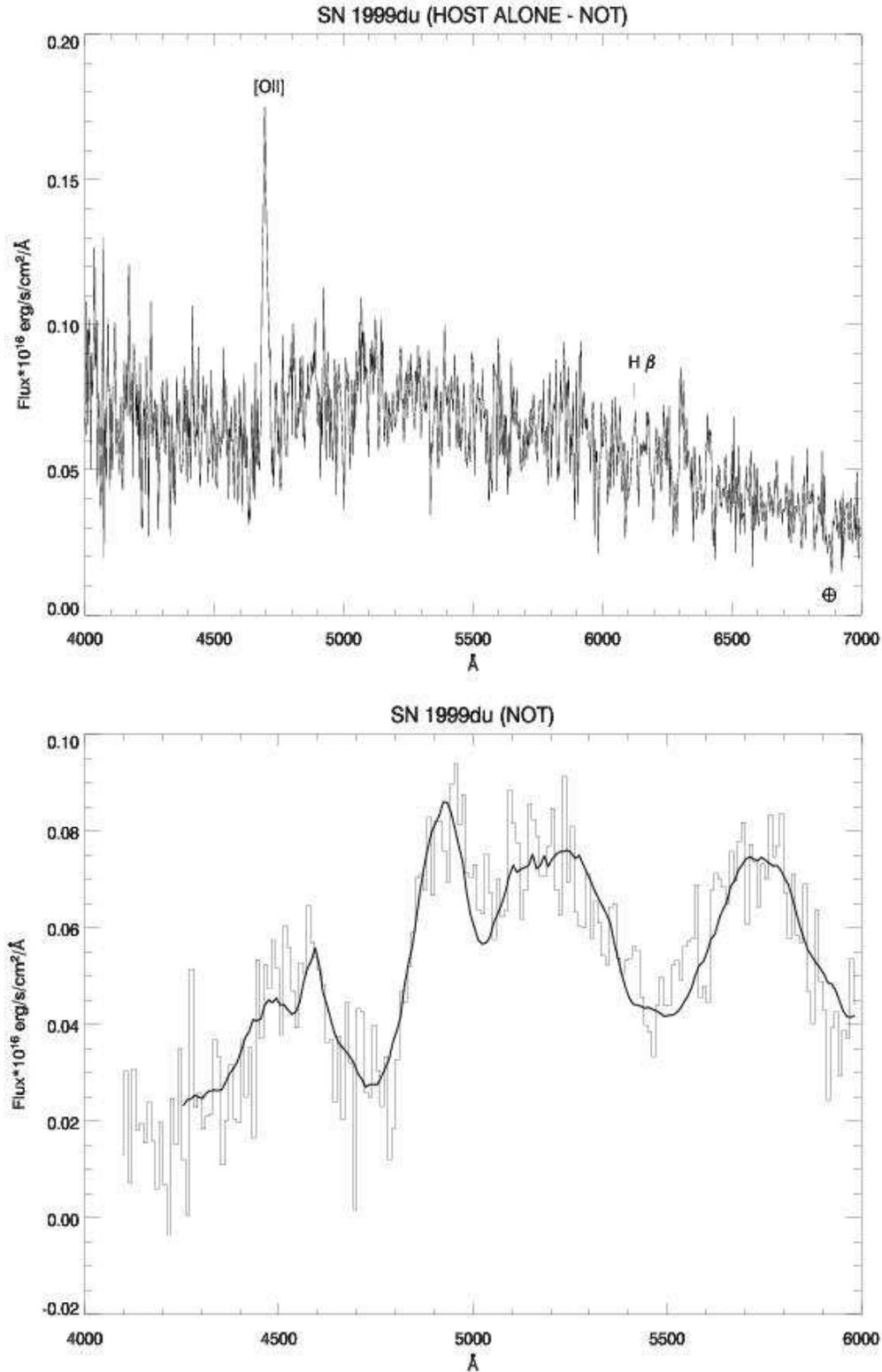} }
\caption{Top: NOT spectrum of SN~1999du host alone. 
  Due to the presence of the second order in the NOT spectra, the
  signal beyond $\sim 6000$ \AA$\ $ is distorted and can not be used
  in the identification fitting procedure. It however shows the
  presence of galaxy lines ($H_\beta$ in this case) useful for
  redshift determination. Bottom: Rebinned galaxy subtracted spectrum
  of SN~1999du with the best-fit template (SN~1992A +6 days)
  overlapped.  The spectrum is shown in the observer frame and is not
  corrected for atmospheric absorptions or galactic line subtraction
  residuals. The spectral range of the bottom panel does not match the
  one of the top panel.}
\label{fig7}
\end{figure*}

\begin{figure*}
 \resizebox{14cm}{!}{\includegraphics{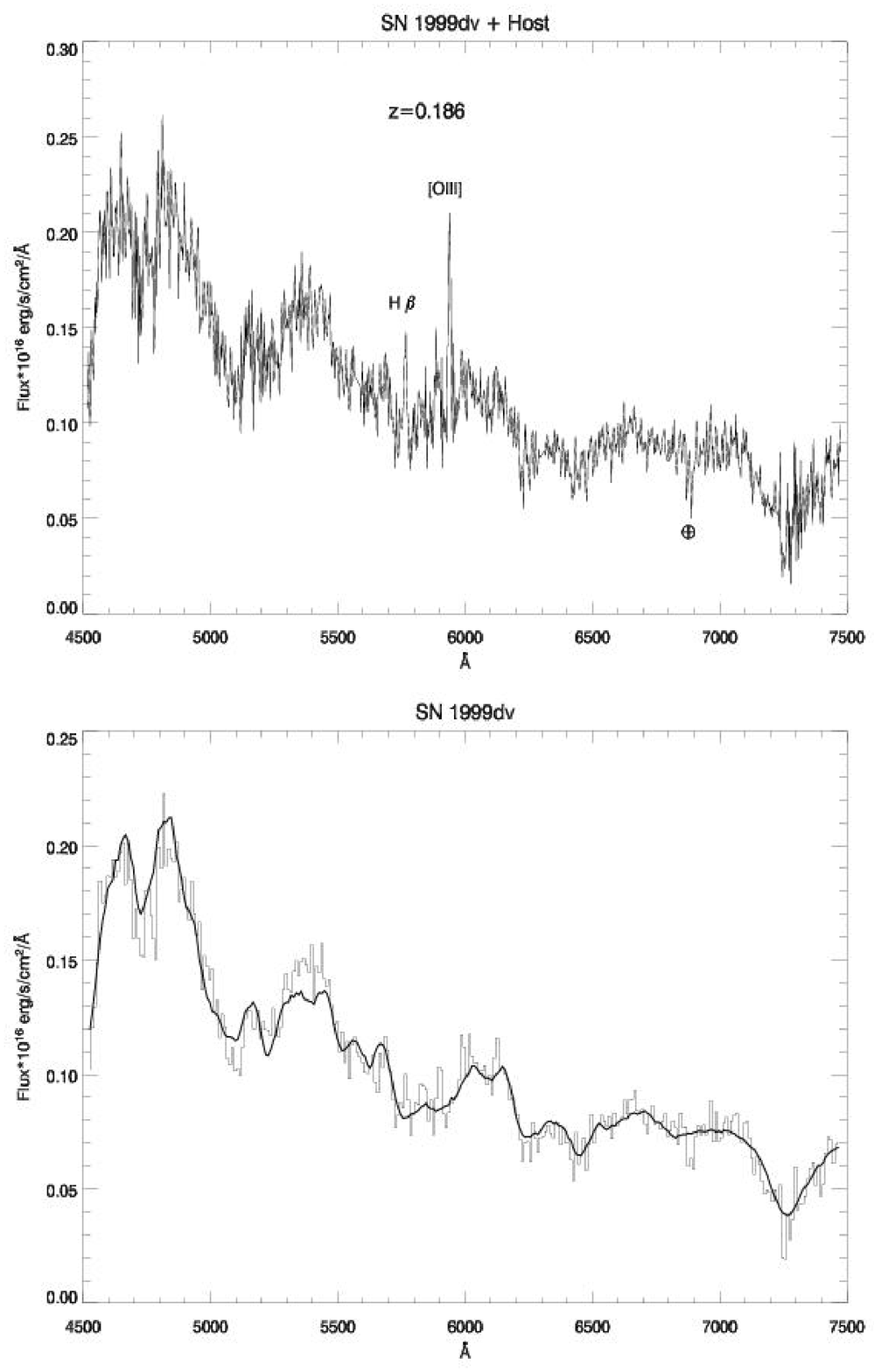} }
\caption{WHT spectrum of SN~1999dv+host. 
  Bottom: Rebinned galaxy-subtracted spectrum of SN~1999dv with the
  best-fit template (SN~2003du -7 days) overlapped. The spectrum is
  shown in the observer frame and is not corrected for atmospheric
  absorptions and galactic line subtraction residuals.}
\label{fig8}
\end{figure*}

\begin{figure*}
 \resizebox{14cm}{!}{\includegraphics{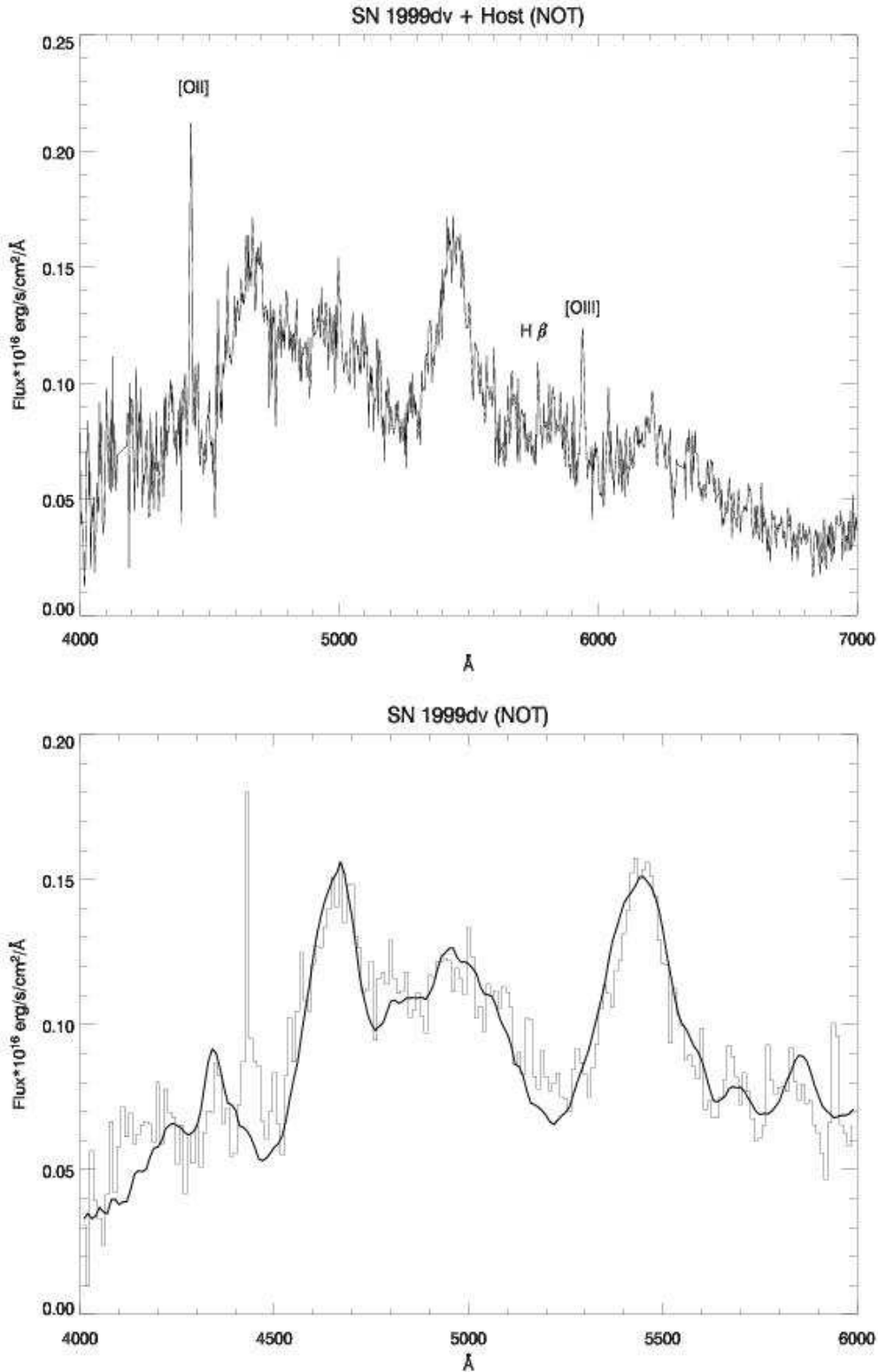} }
\caption{Top: NOT spectrum of SN~1999dv+host. 
  Due to the presence of the second order in the NOT spectra, the
  signal beyond $\sim 6000$ \AA$\ $ is distorted and can not be used
  in the identification fitting procedure. It however shows the
  presence of galaxy lines useful for redshift determination. Bottom:
  Rebinned galaxy-subtracted spectrum of SN~1999dv with the best-fit
  template (SN~1992A +9 days) overlapped.  The spectrum is shown in
  the observer frame and is not corrected for atmospheric absorptions
  or galactic line subtraction residuals. The spectral range of the
  bottom panel does not match the one of the top panel.}
\label{fig9}
\end{figure*}

\begin{figure*}
\resizebox{14cm}{!}{\includegraphics{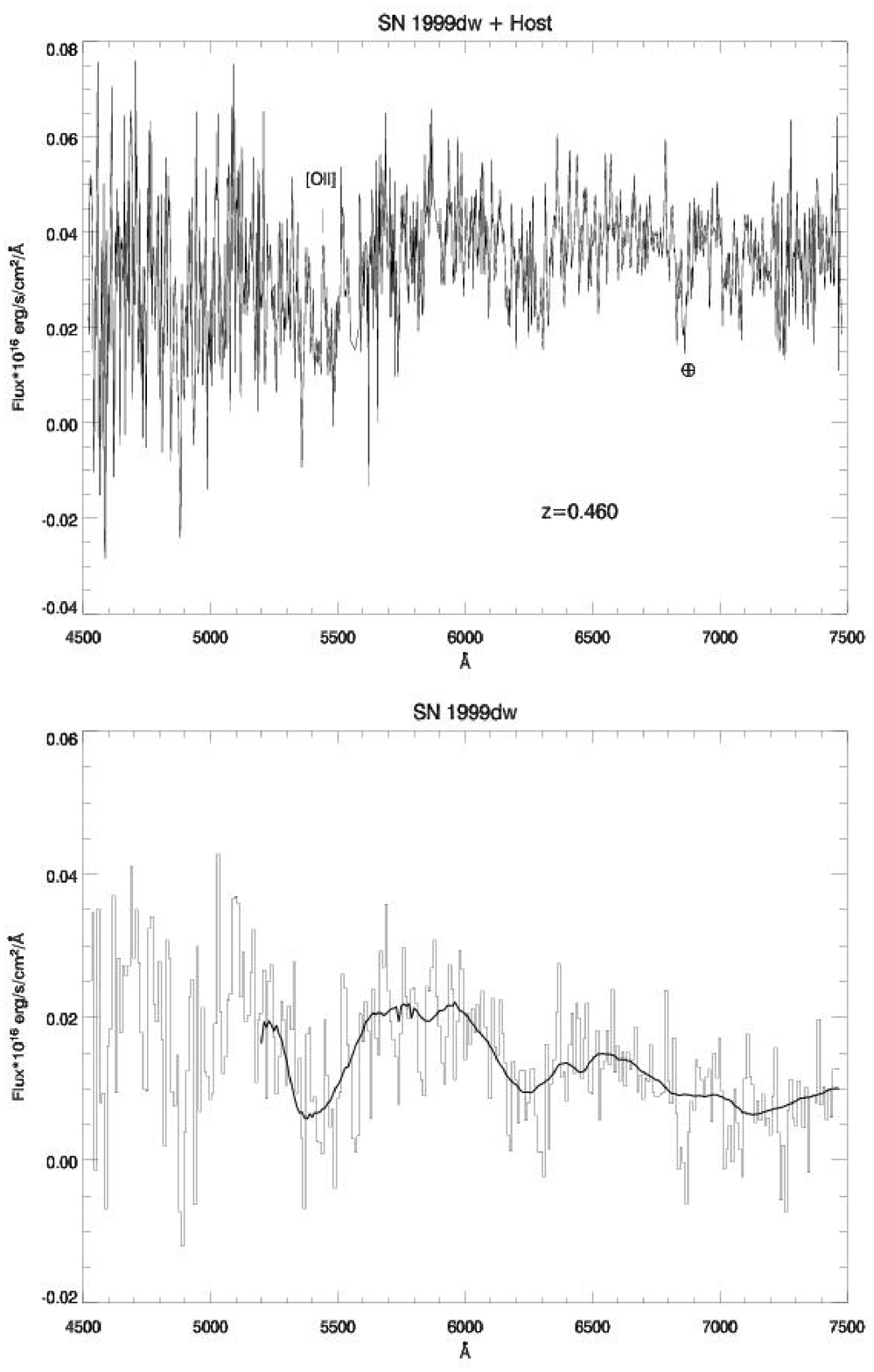} }
\caption{Top: WHT spectrum of SN~1999dw+host. 
  Bottom: Rebinned galaxy-subtracted spectrum of SN~1999dw with the
  best-fit template (SN~1999ee -4 days) overlapped. The spectrum is
  shown in the observer frame and is not corrected for atmospheric
  absorptions or galactic line subtraction residuals.}
\label{fig10}
\end{figure*}

\begin{figure*}
 \resizebox{14cm}{!}{\includegraphics{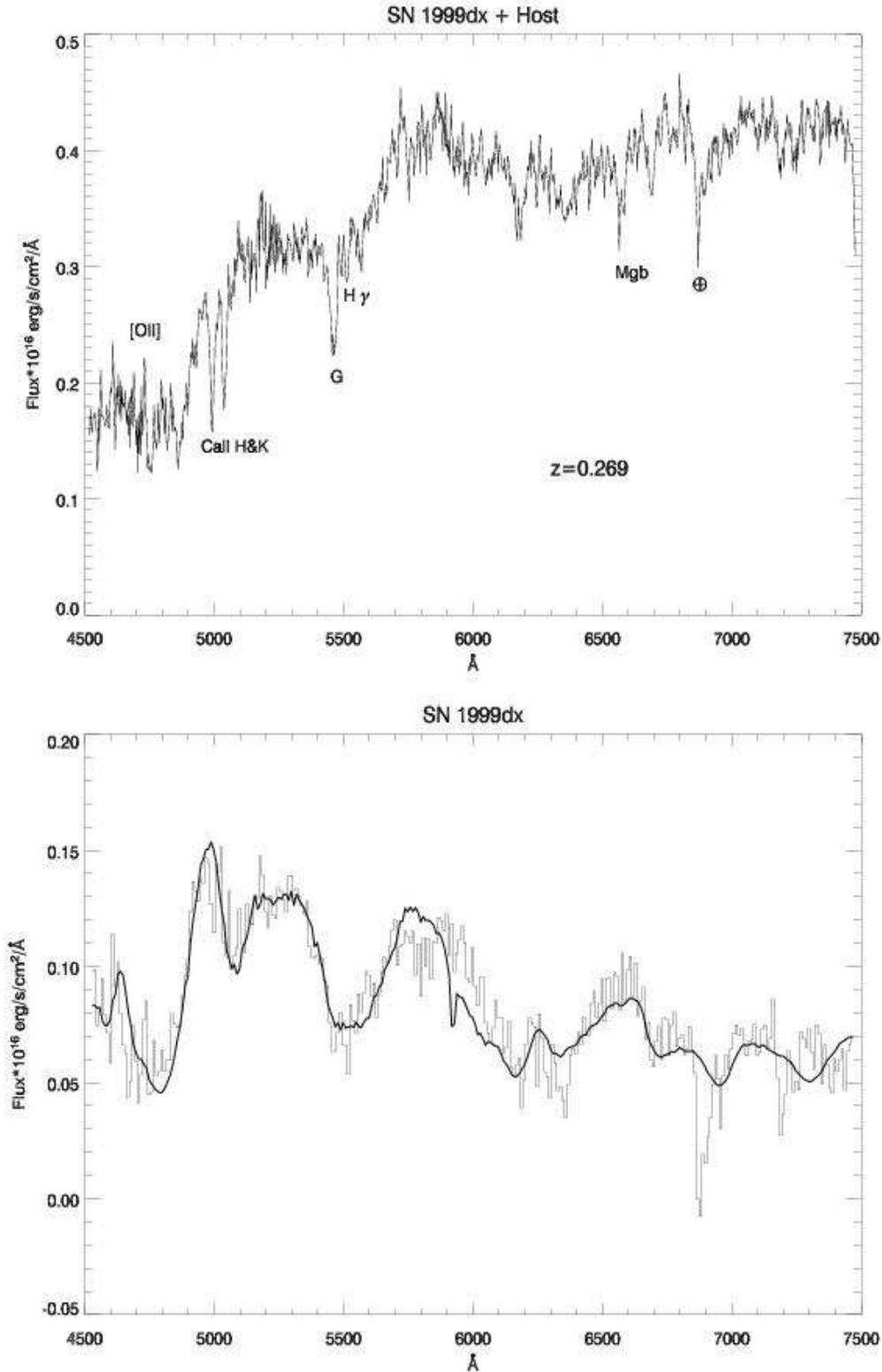} }
\caption{Top: WHT spectrum of SN~1999dx+host. 
  Bottom: Rebinned galaxy-subtracted spectrum of SN~1999dx with the
  best-fit template (SN~1992A +5 days) overlapped. 
The spectrum is shown in the observer frame and is not corrected for atmospheric
absorptions or galactic line subtraction residuals. 
Note the
  presence of strong atmospheric absorptions and a narrow absorption-like
  feature at 5900 \AA$\ $ in the SN~1992A +5 days template.}
\label{fig11}
\end{figure*}

\begin{figure*}
 \resizebox{14cm}{!}{\includegraphics{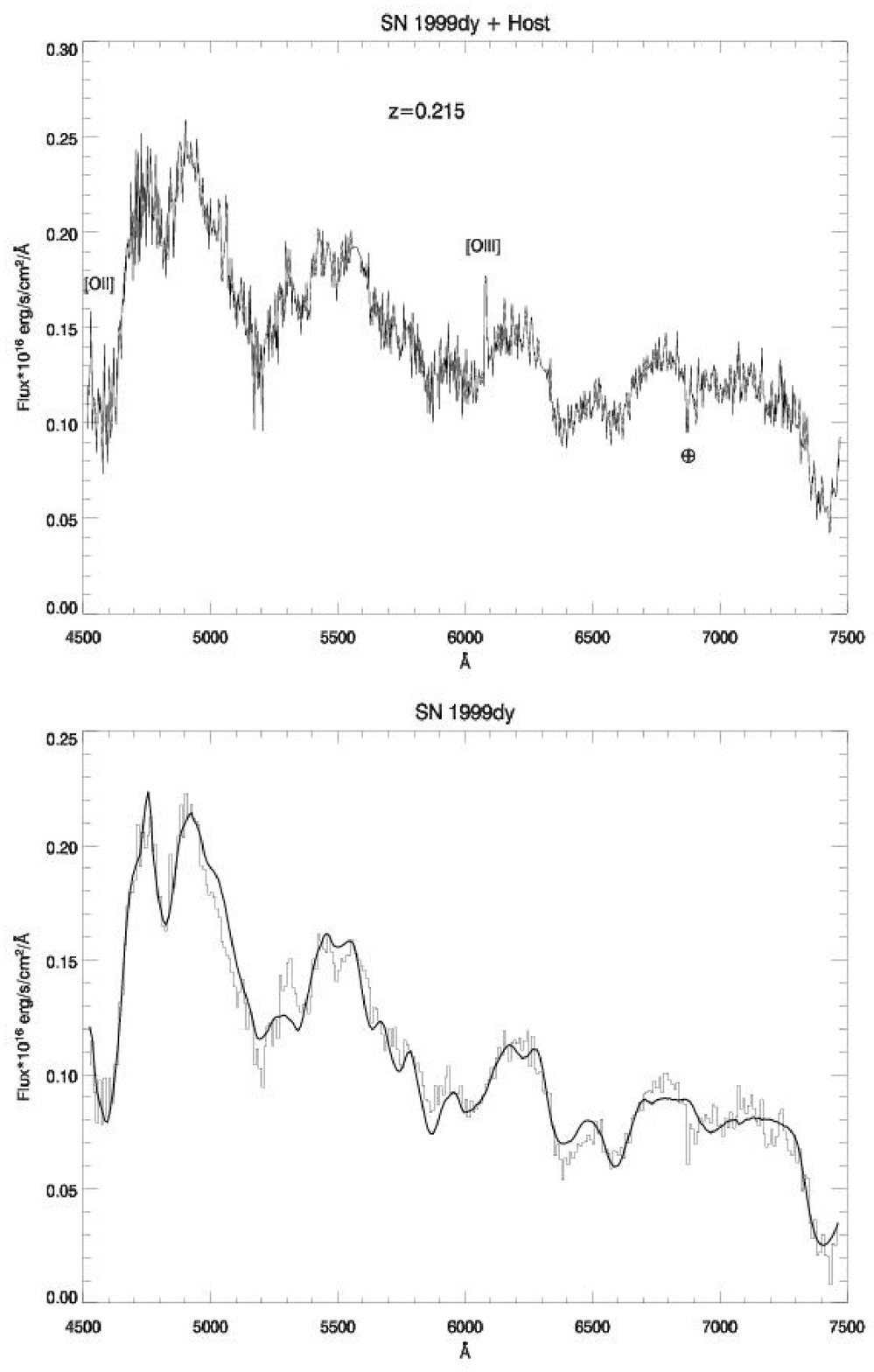} }
\caption{Top: WHT spectrum of SN~1999dy+host. 
  Bottom: Rebinned galaxy-subtracted spectrum of SN~1999dy with the
  best-fit template (SN~1996X 0 day) overlapped. The spectrum is
  shown in the observer frame and is not corrected for atmospheric
  absorptions or galactic line subtraction residuals.}
\label{fig12}
\end{figure*}

\begin{figure*}
\resizebox{14cm}{!}{\includegraphics{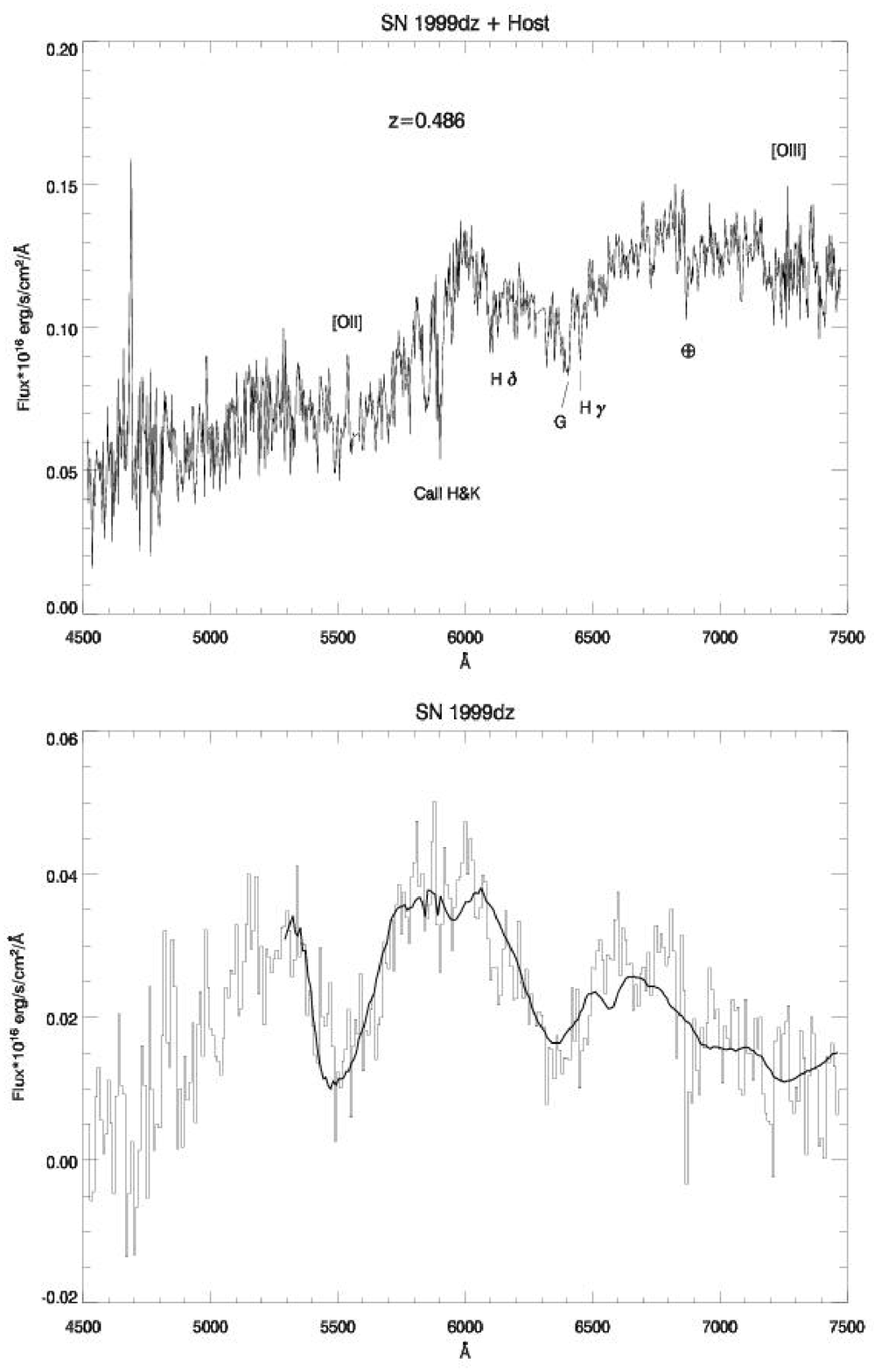} }
\caption{Top: WHT spectrum of SN~1999dz+host. 
  Bottom: Rebinned galaxy-subtracted spectrum of SN~1999dz with the
  best-fit template (SN~1999ee -4 days) overlapped. The spectrum is
  shown in the observer frame and is not corrected for atmospheric
  absorptions or galactic line subtraction residuals.}
\label{fig13}
\end{figure*}

\begin{figure*}
\resizebox{14cm}{!}{\includegraphics{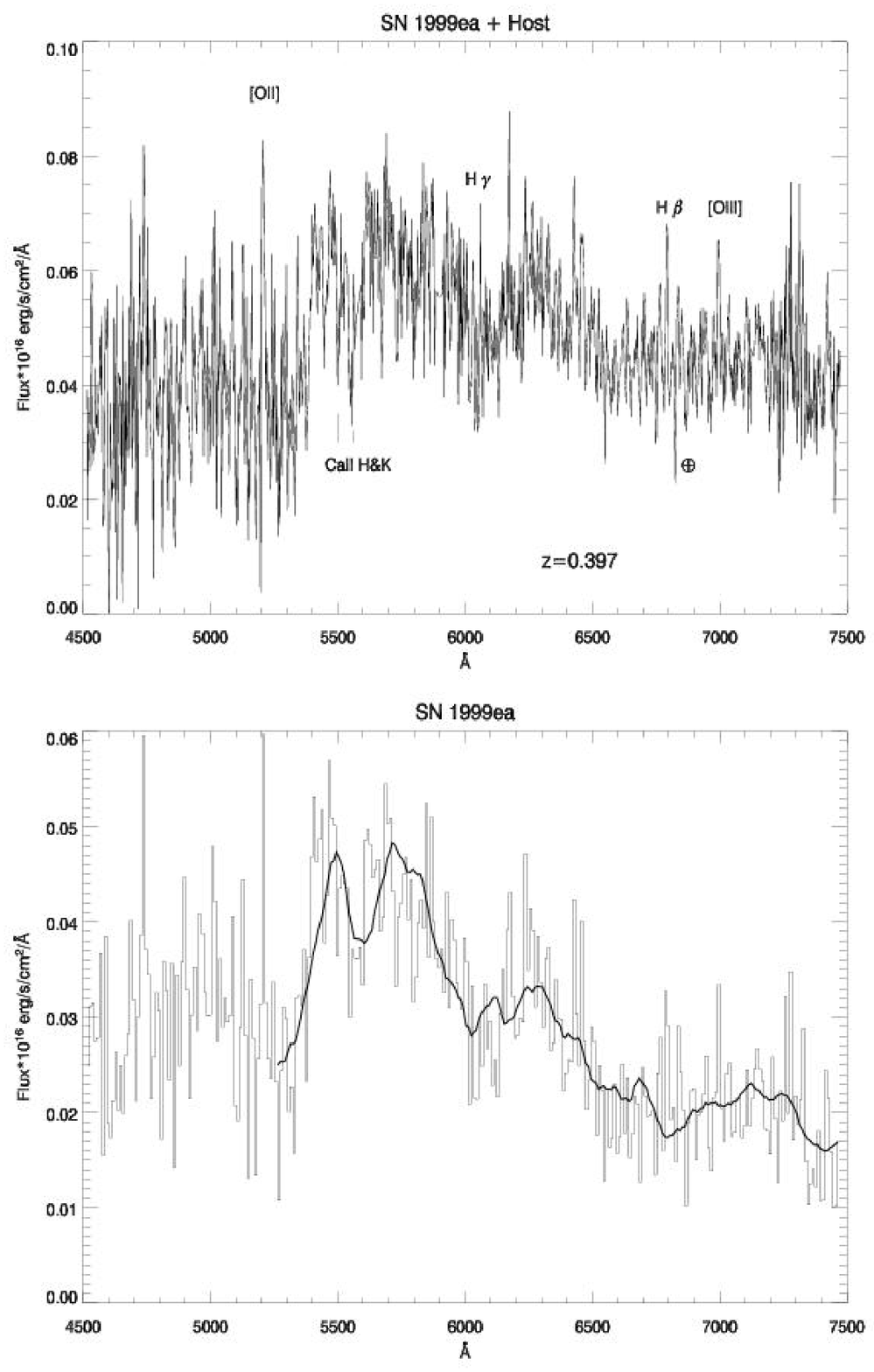} }
\caption{Top: WHT spectrum of SN~1999ea+host. 
  Bottom: Rebinned galaxy-subtracted spectrum of SN~1999ea with the
  best-fit template (SN~1994D -8 days) overlapped. The spectrum is
  shown in the observer frame and is not corrected for atmospheric
  absorptions or galactic line subtraction residuals.}
\label{fig14}
\end{figure*}

\begin{figure*}
 \resizebox{14cm}{!}{\includegraphics{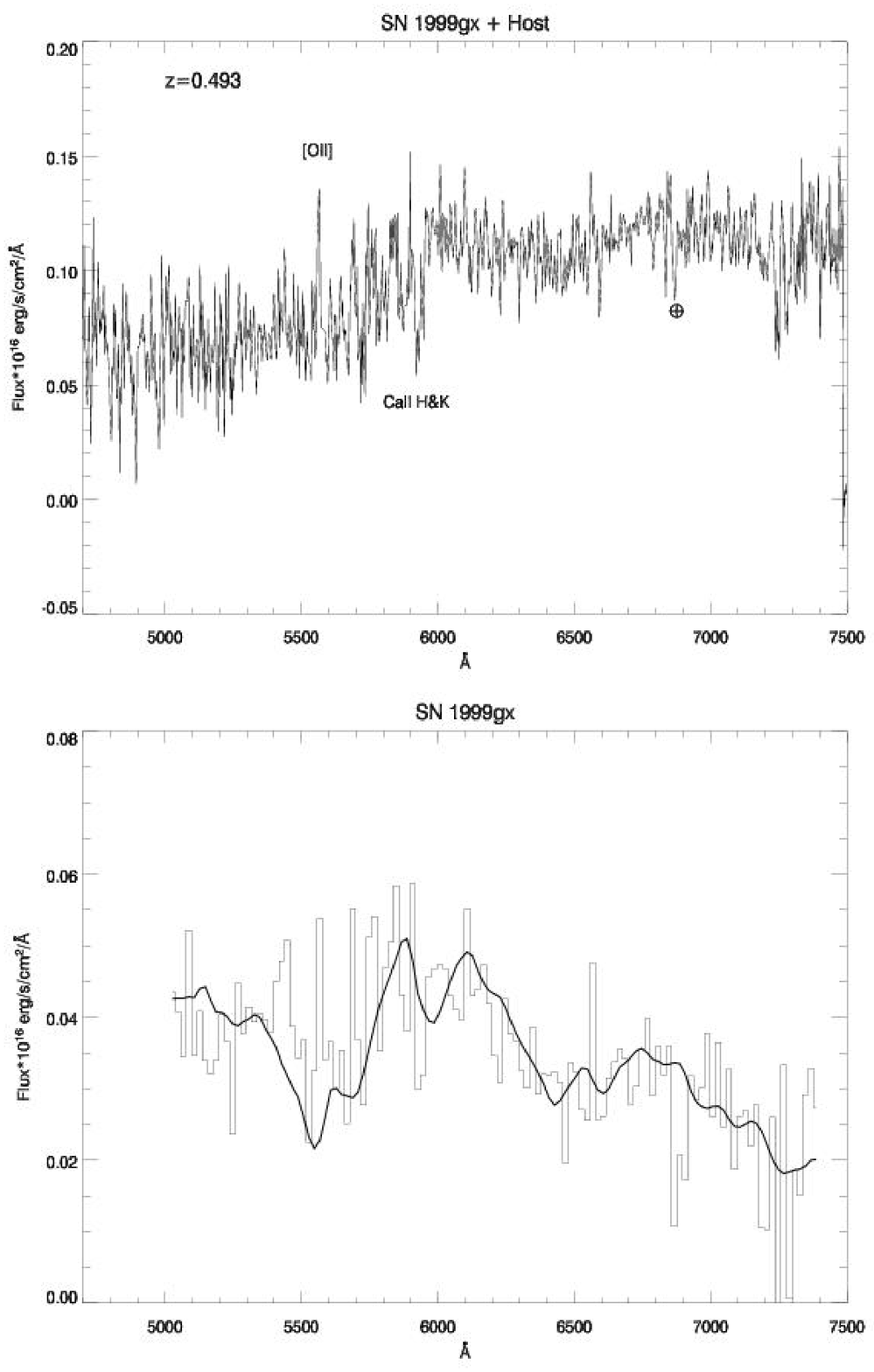} }
\caption{Top: WHT spectrum of SN~1999gx+host. 
  Bottom: Rebinned galaxy-subtracted spectrum of SN~1999gx with the
  best-fit template (SN~1994D -4 days) overlapped. The spectrum is
  shown in the observer frame and is not corrected for atmospheric
  absorptions or galactic line subtraction residuals.}
\label{fig15}
\end{figure*}

We present in the lower panels of Fig. \ref{fig2} to Fig.
\ref{fig15} the output spectrum for the 10+2 confirmed supernovae
along with the best-fit template overlapped on top of each spectrum.
Spectra have been rebinned on 10\AA$\ $ bins for visual convenience.
Top panels show the corresponding host + SN spectra, except for
SN~1999dr and the NOT spectrum of SN~1999du for which separate
extraction of the host galaxy was possible (the galaxy alone is then
shown).  All spectra are presented in the observer frame. Residual sky
lines have been removed 'by hand', but atmospheric absorptions and
galaxy emission/absorption lines are left.  Lines used for redshift
determination are labeled.  The symbols $\oplus$ indicate atmospheric
absorptions.

\begin{figure*}
\resizebox{14cm}{!}{\includegraphics{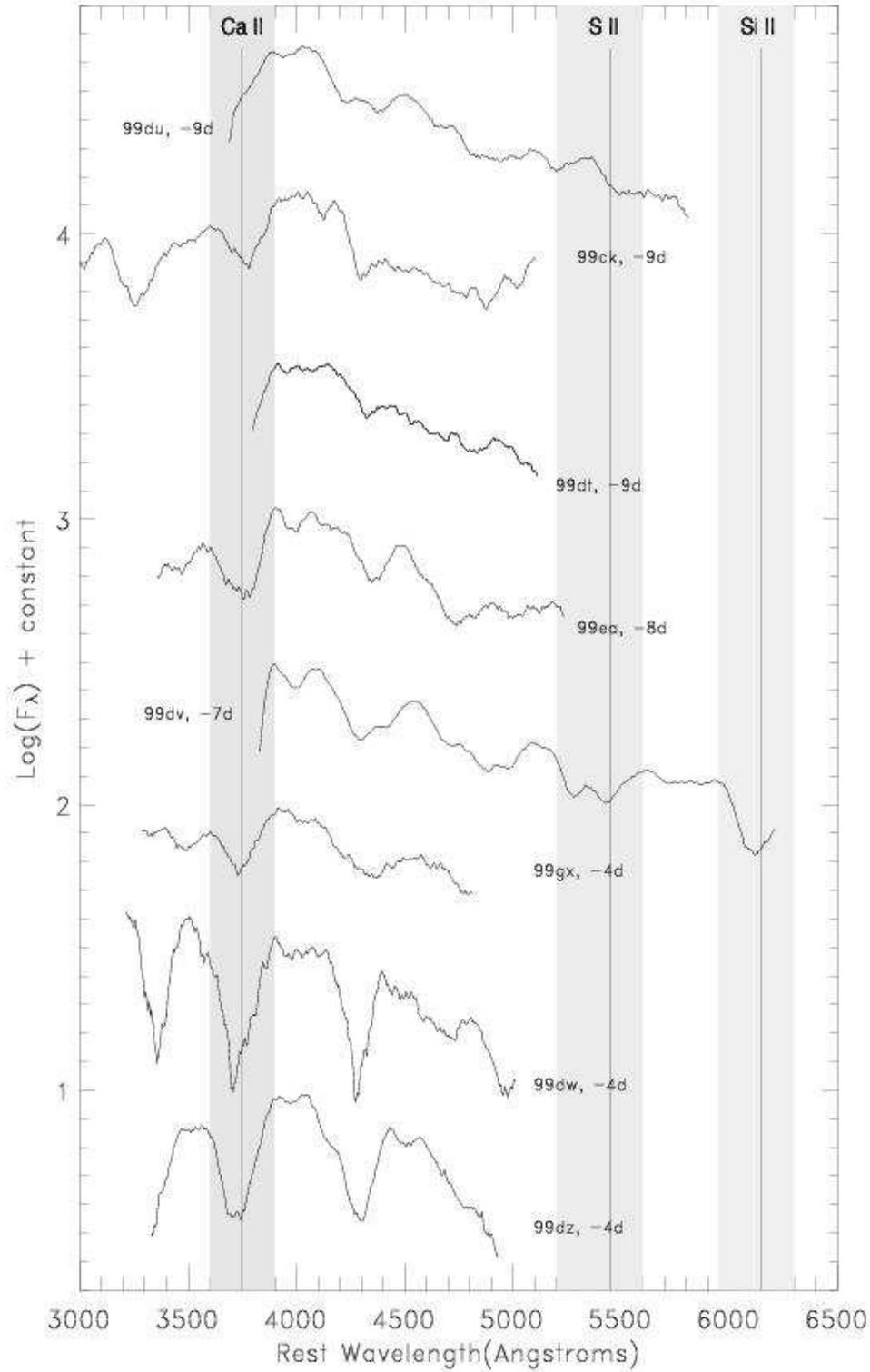} }
\caption{Heavily smoothed spectra of the eight
  pre-maximum \Ia of the intermediate redshift sample ordered in a
  sequence of increasing phase (logarithmic scale). Spectra are
  shifted by an arbitrary amount for visual convenience, and are shown
  into the restframe.  Atmospheric absorptions and galaxy line
  subtractions have been removed before smoothing.  Grey vertical
  bands show the CaII, SII and SiII features found in normal \Ia.
  Solid vertical lines show the positions of CaII at 3945 \AA, SII at
  5640 \AA$\ $ and SiII at 6355 \AA$\ $, blueshifted by 15000 km/s
  (CaII) and 10000 km/s (SII, SiII).  These values are typical of
  'normal' \Ia at maximum and are shown as a guide to the eye.}
\label{fig16}
\end{figure*}

\begin{figure*}
 \resizebox{14cm}{!}{\includegraphics{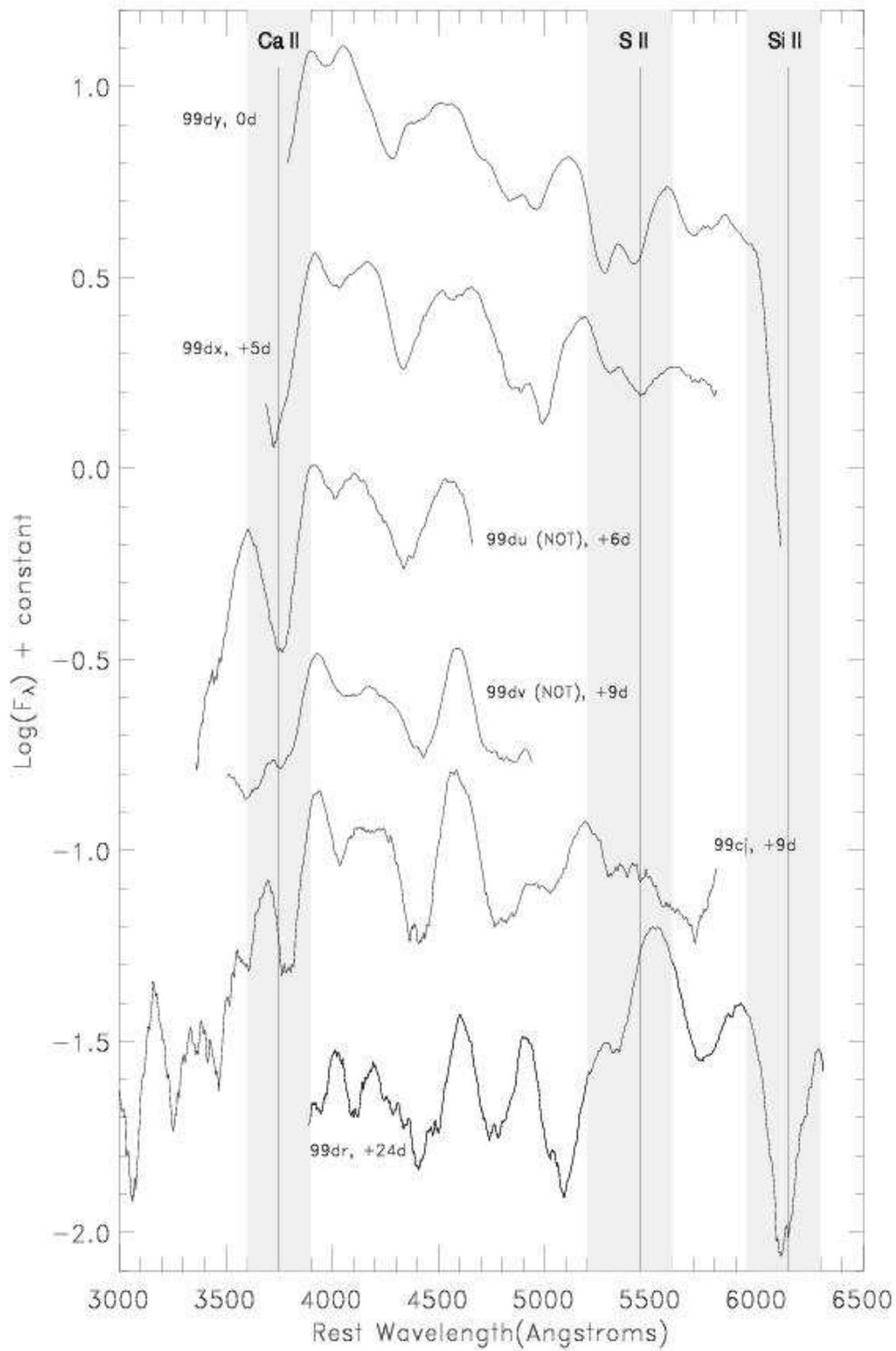} }
\caption{ Same as Fig. 16 for the six past-maximum 
  \Ia (including the two NOT spectra) of the intermediate redshift sample.}
\label{fig17}
\end{figure*}

In Fig. \ref{fig16} and \ref{fig17}, we present the 12 SN classified
in order of increasing spectroscopic phase. Figure \ref{fig16} is for
pre-maximum spectra, whereas Fig. \ref{fig17} shows past-maximum
spectra. All spectra are in the restframe. Residual lines resulting
from imperfect galaxy subtraction, as well as atmospheric absorption
lines have been removed. A polynomial least-squares fit using a
Savitsky-Golay smoothing filter of degree 2 with a window width of 60
data points \citep{Press86} has been applied to the data. This filter
is designed to preserve higher moments within the data and is well
suited for supernova spectra.  For visual convenience, the spectra are
shifted in flux from one another by an arbitrary amount.  The
three gray vertical bands show the CaII, SII and SiII spectral
features expected to be found in 'normal' \Ia.  In addition, three
vertical solid lines indicate the positions of CaII at 3945 \AA, SII
at 5640 \AA$\ $ and SiII at 6355 \AA$\ $, blueshifted by 15000 km/s
(CaII) and 10000 km/s (SII, SiII).  These values are typical of
'normal' \Ia at maximum \citep{Benetti04} and are shown as a guide to
the eye.

Early spectra of 'normal' \Ia are usually characterized by the
presence of intermediate mass elements such as Si, S and Ca. As the
spectrum evolves, they are progressively replaced by features due to
iron-peak ions.  A strong absorption feature of SiII visible at $6150$
\AA$\ $ in spectra around maximum light is probably the most
discriminant feature against other SN types
\citep{Wheeler90,Filippenko97,Leibundgut00}.  However, this feature is
only present in two of our spectra (SN~1999dr and SN~1999dv) whereas
for all other SN it is redshifted beyond the upper limit of $7500$
\AA$\ $ even at moderate redshift.  Other characteristic features we
base our identification on include:

$\bullet$ Presence of a W-shape visible around 5500 \AA$\ $ (restframe)
due to SII $5640$ \AA$\ $

$\bullet$ Presence of an absorption feature due to SiII $4130$ \AA.
This feature can be weak in over-luminous supernovae of SN~1991T type
prior to maximum \citep{Filippenko92,Ruiz92}. It is on the contrary
abnormally deep and distorted in under-luminous supernovae like SN
1991bg due to the presence of TiII absorption in the same wavelength
region \citep{Li01b,Garnavich04}.  Note that for Type Ib/c supernovae,
this feature presents a plateau-shape longward of $4100$ \AA$\ $ up to
about $4400$ \AA$\ $ \citep{Filippenko97,Branch02} similarly to its
appearance in SN~1991bg-like SNe
\citep{Leibundgut93,Turatto96,Mazzali97,Li01b}.

$\bullet$ Presence of a strong absorption due to CaII $3945$ \AA.
This feature is not characteristic of \Ia as it exists as well in
other SN types and under-luminous \Ia. However, it is weak in
SN~1991T-like objects prior to maximum light \citep{Li01b}.

In the following, we briefly describe the extracted spectrum for each
supernova.
 
{\bf SN 1999cj:} Due to the use of the blue arm, part of the restframe
UV is accessible for the two April \Ia. SN~1999cj (Fig. \ref{fig2}) is
at $z_{h}=0.362$ and exhibits features characteristics of a \Ia well
after maximum. The best-fit is obtained with SN~1992A, 9 days past
maximum \citep{Kirshner93}. In particular, suppression of the peak
redward of the SiII absorption around $4130$ \AA$\ $ (restframe) is
clearly seen.

{\bf SN 1999ck:} An early \Ia at $z_h=0.432$ with a UV part from $\sim
3000$ \AA$\ $ restframe (Fig. \ref{fig3}). Bumps and troughs are
clearly seen in this part of the spectrum. Best fit obtained with
SN~1994D -9 days \citep{Patat96}.

{\bf SN 1999dr:} This is a \Ia at $z_h=0.178$. The host spectrum has
been extracted separately from the SN and used as the galaxy component
of the fitting model. The supernova spectrum is consistent with a \Ia
(best fit with SN~1994D) at 24 days after maximum light. As spectra
have been obtained on September 12th, that places the date of maximum
around August 19, 1999 so that the supernova was starting rising when
the reference image was taken on August 9.  The discovery image was
obtained 23 days after the reference. At that date, the flux
difference between the reference and the discovery image still permits
it to be detected as a light increasing supernova.  Even though our
detection strategy has been designed to select rising SNe, decreasing
SNe can also be found.  The aspect of the spectrum (Fig. \ref{fig4})
is typical of a supernova in the photospheric phase a few weeks after
maximum, as Fe lines develop and create large troughs in the spectrum.
This is clearly seen around $6000$ \AA$\ $ ($\approx 5100$ \AA$\ $
restframe) where the large dip is due to a blend of FeII absorption
lines. The bump at $\approx 6500$ \AA$\ $ replaces the W-shape due to
SII seen in spectra around maximum or earlier.

{\bf SN 1999dt:} This is the fourth most distant SN of our sample
($z_h=0.437$). The spectrum is strongly dominated by the host galaxy
and the host-subtracted spectrum presented has thus a rather poor S/N
(see Fig. \ref{fig5}). The best-fit is for an early SN template
(SN~1994D -9 days).  The three best-fit solutions (in terms of
$\chi^2$) all give a 'normal' \Ia but the third one differs in phase
by 7-8 days with respect to the two others: the best-fit solution is
for -9 days while the third one yields a -1 day spectrum template.
This 'phase instability' is not surprising as the spectrum is strongly
dominated by the host.  Unfortunately, as most of the candidates are
detected within 1.5 arcsec of the host center comparable to the
seeing, it is not possible to extract separately both components.

{\bf SN 1999du:} SN 1999du has been observed at WHT on September 12
and at NOT on October 2. Fig. \ref{fig6} and \ref{fig7} show the
corresponding spectra.  For the NOT spectrum (Fig. \ref{fig7}), the
wavelength upper limit adopted for the fit is 6000 \AA. Indeed, the
signal redward of this value is polluted by the presence of the second
order spectrum and is practically useless for our fitting purposes.
The same remark applies for SN~1999dv.  The October 2 spectrum is best
fitted by a \Ia (SN~1992A) 6 days after maximum whereas a \Ia
(SN~1999ee \citep{Hamuy02}) at -9 days matches the September 12
spectrum. Taking the September 12 determination at face value, this
places the date of maximum light on September 21, while the NOT
determination leads to September 26 for maximum. This is marginally
consistent given the adopted $\pm 3$ days of uncertainty in the
spectrum phase determination. Correcting for the time dilation effect
expected in an expanding universe \citep{Goldhaber01, Riess97,
  Foley05} gives the date of maximum light on September 23 and
September 24 for the WHT and NOT spectra, respectively, a fully
consistent result.  The host redshift determination agrees for the two
spectra, leading to $z_h=0.260$.  Note that for the NOT spectrum, the
host galaxy has been extracted separately and used to construct the
best-fit model, as for SN~1999dr. With this procedure, the fit is
performed against the redshift of the supernova alone, not against the
redshift of the supernova+galaxy template model, as for the other WHT
spectra. This leads to a difference in $z_f$ and $z_h$ of 0.01 (see
Table \ref{tab2}), the largest difference observed for our sample
(along with SN~1999dy). This value can be considered as an upper limit
on the precision on the best-fit redshift $z_f$. If we do not use the
observed galaxy and rather use a template galaxy for the fitting
model, this redshift discrepancy disappears (we then find
$z_{f}=0.263$).  The best-fit for the WHT spectrum is obtained for a
reddened starburst galaxy with E(B-V)=0.4, labeled as 'Stb4' in our
database \citep{Kinney96}. The same solution is obtained for the NOT
spectrum when a galaxy template is used for the fit rather than the
observed spectrum.  This solution indicates a dusty star-forming
galaxy.  Indeed, strong emission lines are seen in the full spectrum.
Note that excluding these lines from the fit leads to a solution with
the same SN template (SN~1992A +6 days) with a slightly lower $\chi^2$.

{\bf SN 1999dv:} As for SN 1999du, both WHT and NOT spectra have been
obtained for SN 1999dv and are shown on Fig. \ref{fig8} and
\ref{fig9}. The WHT spectrum is well fitted by a \Ia - SN~2003du
\citep{Anupama05} at -7 days, and is only weakly contaminated by its
host. The host redshift is $z_h=0.186$, fully consistent with the
determination from the NOT host spectrum. This latter spectrum is best
fitted by a late \Ia (SN~1993A) template at +9 days. Using the WHT
phase, the maximum date falls on September 20, whereas the NOT phase
determination leads to a maximum on September 23. These dates of
maximum are consistent within the $\pm 3$ day uncertainty in the phase
determination.  Note that the phase difference found between the two
observations (16 days) is similar within one day with the difference
obtained for SN~1999du (15 days) but is short, by 3 to 4 days, of the
real number of days elapsed between the WHT and NOT observations.
Correcting for time dilation leads to $\sim 19$ days difference in the
observer frame. As for SN~1999du, this is fully consistent with the
actual observing dates.

{\bf SN 1999dw:} The third most distant \Ia of our sample at
$z_h=0.460$. The spectrum is shown in Fig. \ref{fig10}. It is well
fitted by an early template (SN~1999ee $ -4$ days) and exhibits
features of a \Ia at this date. The CaII absorption feature
corresponding to $\lambda \approx 3945$ \AA$\ $ restframe is clearly
visible. Not however that the spectrum seems slightly bluer than other
\Ia at a comparable phase.
 
{\bf SN 1999dx:} This is a heavily buried \Ia at $z_h=0.269$ with a
host contribution of more 75 \% (Fig. \ref{fig11}). The best-fit is
obtained for a +5 days \Ia template (SN~1992A).  The fit is poor in
the range 5800 - 6200 \AA$\ $ as the template falls below the observed
spectrum.  Interestingly, this high signal-to-noise supernova yields
the poorest best-fit reduced $\chi^2$ of our sample ($\geq 1.5$).
Indeed, the higher the quality of the spectrum (the lower the noise),
the more the $\chi^2$ value reflects discrepancies with the underlying
model. This illustrates how our fitting procedure depends on the
quality of the underlying model. As explained in \S 4.2, this is
directly connected to the phase sampling and the number of the
spectral SN templates in our database.
 
{\bf SN 1999dy:} A typical \Ia at maximum with a high signal-to-noise
ratio.  The redshift is $z_h=0.215$ (from galaxy lines, $z_f$ being
significantly lower at around $z_f=0.202$) . The best fit is obtained
with SN~1996X \citep{Salvo01} and is very good except around $\lambda
\sim 5300$ \AA$\ $ where an emission feature is present.  Around
$\lambda \sim 6300$ \AA, the spectrum falls below the fit, maybe due
to an uncorrected $O_2$ atmospheric absorption feature around this
wavelength (Fig. \ref{fig12}).
 
{\bf SN 1999dz:} This second most distant \Ia ($z_h=0.486$) is
presented in Fig. \ref{fig13}. The fitting procedure yields a large
fraction of host galaxy and an early \Ia (SN~1999ee $-4$ days).  The
phase is rather stable when one considers solutions of immediately
higher $\chi^2$.  Due to its fairly high redshift, features blueward
to $\lambda \sim 4000$ \AA$\ $(restframe) are visible. Note that the
two peaks in the fit spectrum around $6400 - 6900$ \AA$\ $ appear
blueshifted with respect to the peaks in the observed spectrum.  It is
difficult to determine whether this discrepancy is due to an intrinsic
feature of the supernova. As the galaxy contribution to the fit model
is almost 80\%, it might be due to galaxy subtraction residuals.

{\bf SN 1999ea:} This is a \Ia very similar to SN 1999dz at a lower
redshift ($z_h=0.397$). The best fit is obtained for a \Ia (SN~1994D)
at -8 days, again a rather stable phase (Fig. \ref{fig14}).

{\bf SN~1999gx:} The most distant \Ia ($z_h=0.493$) of our sample
(Fig.  \ref{fig15}). The SN signal is rather weak and was not
identified during the first assessment at the telescope. Re-reduction
of the data allowed to extract this distant SN \citep{Balland05} whose
spectrum is best fitted by a normal \Ia template (SN~1994D +6 days).

\subsection{Properties of the SN Ia}

As far as $\chi^2$ values are concerned, all the spectra presented in
this paper are best fitted with templates of normal \Ia as opposed to
peculiar \Ia. This is true even for SN~1999gx, even though $\chi^2$
values are very close (see Table \ref{tab7}).  Given their
signal-to-noise and redshifts, it is possible to compute a few
spectral quantities of interest to further characterize these
supernovae. This is an important study to perform as little is known
about the spectral properties of \Ia at intermediate redshift and the
possible differences with their low-z counterparts. For instance,
finding spectral peculiarities in samples at intermediate or high
redshift would shed light on the evolution of the 'peculiarity rate'
with redshift \citep{Li01b}. This might also give some hint of how
samples used for the determination of cosmological parameters might be
contaminated by peculiar supernovae.
 
\subsubsection{Velocity measurements}

Even \Ia classified as 'Branch normal' do not form an homogeneous
spectral class of objects.  For instance, differences in evolution of
CaII, SII and SiII absorption lines velocities are found among them
\citep{Branch88,Hatano00,Benetti04}.  \Ia such as 2002bo or 1984A for
instance exhibit rather high velocities $v_{SiII}$ or $v_{CaII}$
compared to other supernovae considered as 'Branch normal'
\citep{Benetti04}. This might indicate some degree of peculiarity.
However, as more spectra of nearby supernovae are collected and
studied, there is evidence that the standard classification in
'normal', under- or over-luminous objects is far too simple to
accurately reflect diversity among supernovae.  \citet{Benetti05}
suggest a classification in 'faint' SNe with high-SiII velocity
gradients, 'normal' SNe, also with high-velocity gradients but with
brighter absolute magnitude, and SNe with low velocity gradients.
These three classes are related to the progenitor explosion mechanism.
In this classification, 'Branch-normal' \Ia are found in both the
'faint' and 'normal' classes.

\begin{figure*}
 \resizebox{14cm}{!}{\includegraphics{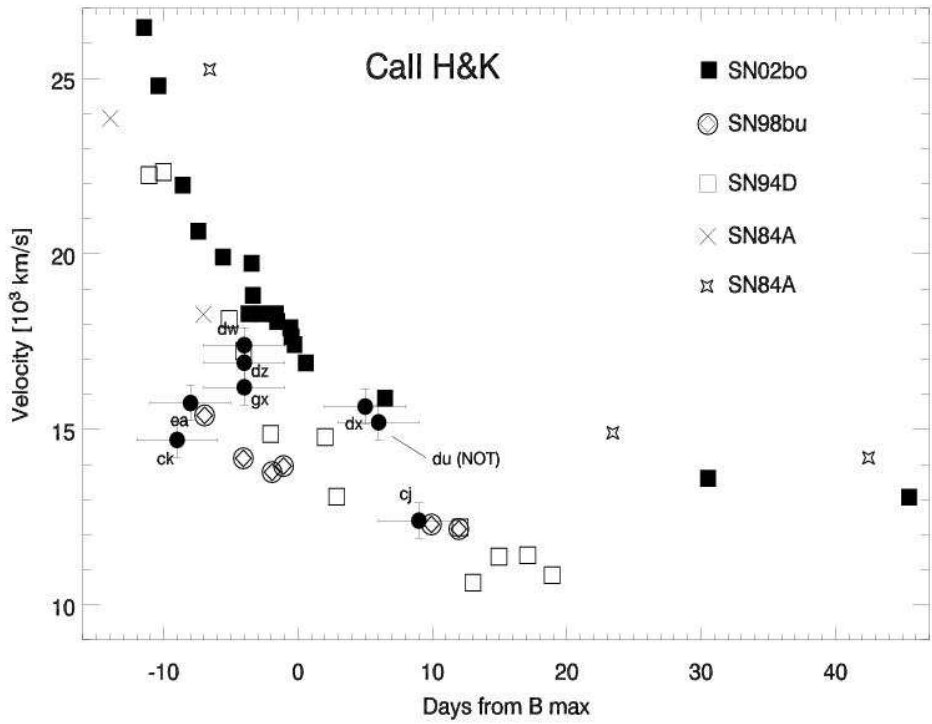} }
\caption{CaII velocity as a function of B-band phase.
  Data are taken from Figure 11 of Benetti \etal (2004). Calcium
  velocities for the \Ia presented in this paper (plotted as filled
  circles) are given in Table 3.}
\label{fig18}
\end{figure*}

Velocity gradient measurements are unfortunately out of reach for our
sample, even for the two \Ia for which we have spectra on two
different dates.  CaII, SII and SiII absorption line velocities can
be however measured in some cases. Results are shown in Table
\ref{tab4}. Velocities have been obtained by fitting the corresponding
line feature by a gaussian profile. Typically, the error is 500 km/s
for Ca and 200 km/s for both SII and SiII velocities.  We compare our
velocity results to the ones of local supernovae Ia by producing
'velocity vs day from B maximum' plots following, e.g.
\citet{Benetti04}. Figure \ref{fig18} shows such a plot for the CaII
H\& K feature. Data points are taken from figure 11 of
\citet{Benetti04} (see references therein).  The date from B-maximum
is estimated from the best-fit phase of Table \ref{tab2} except for 5
\Ia for which g' and r' photometry are available (see discussion
below).

Our different measurements are consistent with values for 'Branch
normal' \Ia. The SiII velocity of SN~1999dr is slightly high, in
between the one for SN~1994D and SN~1984A.  SN~1999dw, SN~1999dz and
SN~1999gx CaII H\& K velocities are close to the ones of SN~1994D,
while SN~1999dx and SN~1999du $v_{CaII}$ are closer to the one of
SN~2002bo. SN~1999ea and SN~1999ck CaII  velocities are low, however
consistent with SN~1998bu velocity.  Despite this diversity, there is
no hint from velocity measurements that any of our supernovae are
peculiar.

\subsubsection{Searching for peculiar SN Ia}

We have used spectral information to search for peculiarities in the
SNe of our sample (search for peculiar features in the spectra,
measurements of absorption line velocities) and found that all our \Ia
are consistent with being 'normal' \Ia. In order to assess the
validity of this conclusion, we also fit our spectra with ${\cal
  SN}$-fit using only the peculiar \Ia spectral templates available in
our database.  We dispose of templates for eight peculiar SNe,
both under and over-luminous with a similar phase sampling as for
'normal' SNe Ia.  We list the results in Table \ref{tab7}.  In columns
3 and 4 the best-fit template and phase obtained when fitting with
peculiar \Ia templates are given. The corresponding reduced $\chi^2$
and number of degrees of freedom (d.o.f.) are presented in columns 5
and 6.  For comparison, we also report the same parameters for
'normal' \Ia templates from Table \ref{tab2} (columns 7 to 9). When
fitting with peculiar templates alone, best-fits are systematically
obtained for over-luminous \Ia, as seen in column 3 (templates for
SN~1991T are from \citet{Mazzali95}; SN~1999aa from
\citet{Garavini04}; SN~2000cx from \citet{Li01c}). From the comparison
of the $\chi^2$ values in columns 5 and 8, it appears that 'normal'
templates are favored over peculiar templates in all cases.  We have
performed a F-test to quantify the significance of the corresponding
$\chi^2$ difference.  Column 10 gives the probability that the two
fits are equivalent. For one \Ia (SN~1999gx), the 'null hypothesis'
that a solution with a peculiar \Ia is equivalent to a solution with a
'normal' one is verified at the $2\sigma$ level. A clear
identification is thus difficult for this object. As seen in Fig.
\ref{fig16}, SN~1999gx exhibits shallower absorptions than other SN at
comparable phases.  This might hint toward a somewhat peculiar
supernova, although note that its CaII velocity is normal. 

\begin{figure*}
 \resizebox{14cm}{!}{\includegraphics{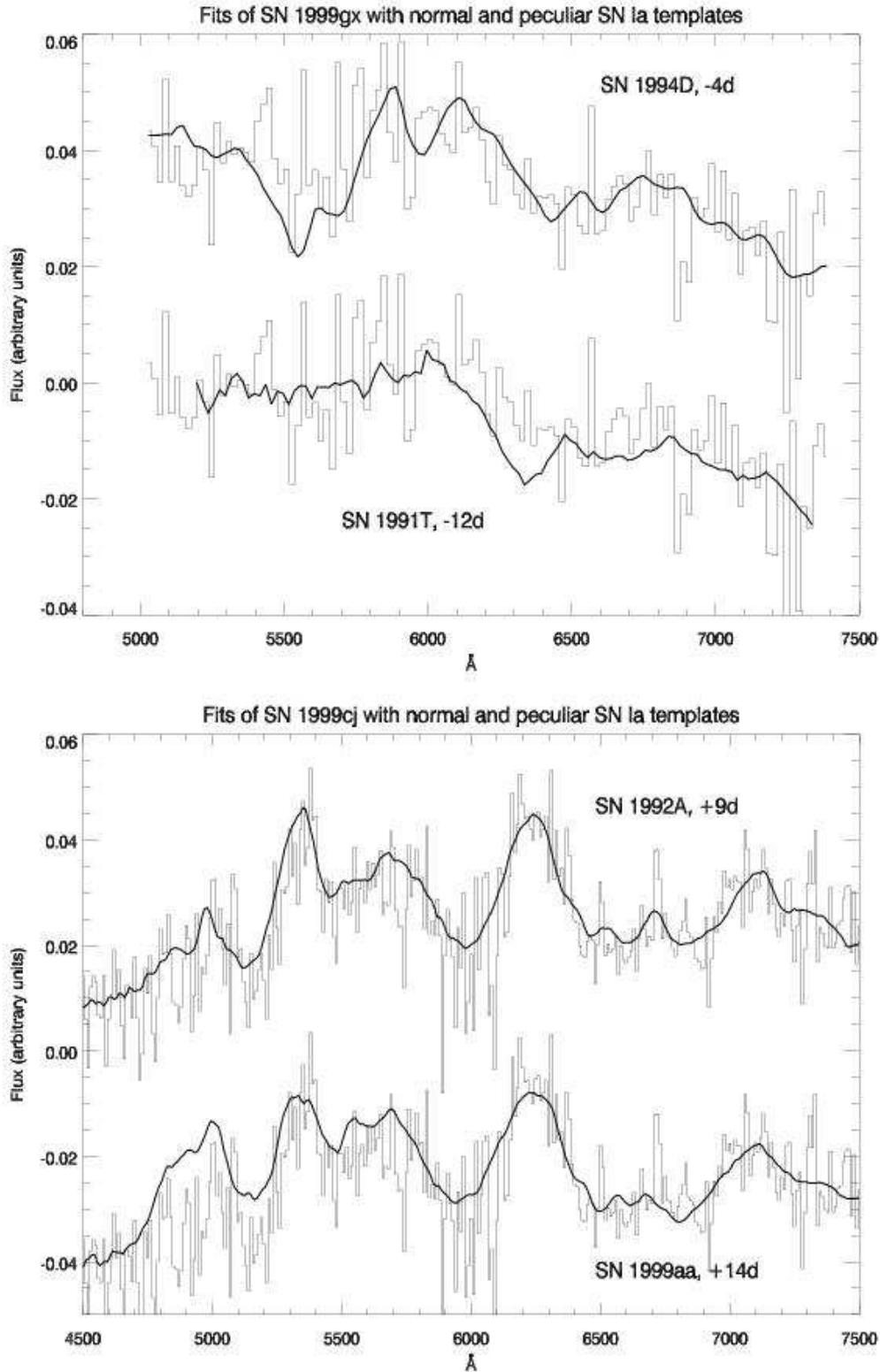} }
\caption{
  Comparison of best fits obtained for 'normal' and peculiar \Ia
  templates for two low S/N ratio supernovae for which identification
  might be questionable. Top: SN~1999gx, and bottom: SN~1999cj. For
  clarity, peculiar solutions have been shifted by an arbitrary amount
  in y-scale. In both cases, visual inspection slightly favors the
  normal solution over the peculiar one.}
\label{fig15bis}
\end{figure*}

Top panel of Fig. \ref{fig15bis} shows the best-fits obtained for
SN~1999gx for a 'normal' and a peculiar solution. For clarity, the
'peculiar' solution (lower curve) has been shifted in flux by an
arbitrary amount. The same graph for SN~1999cj, the supernovae
yielding the second highest F-test probability (57\%) after SN~1999gx,
is shown on the bottom panel. In both cases, visual inspection of Fig.
\ref{fig15bis} slightly favors the 'normal' fit over the peculiar,
although differences tend to be small.

For other \Ia, the 'null hypothesis' is significantly rejected.
For SN~1999dv and SN~1999dy, two supernovae for which host
contamination is weak, the $\chi^2$ increases dramatically, which
translates in very low F-test probabilities.  This strongly supports
the hypothesis that they are 'normal' \Ia.

\subsubsection{\Ia Phases}

\begin{figure*}
 \resizebox{14cm}{!}{\includegraphics{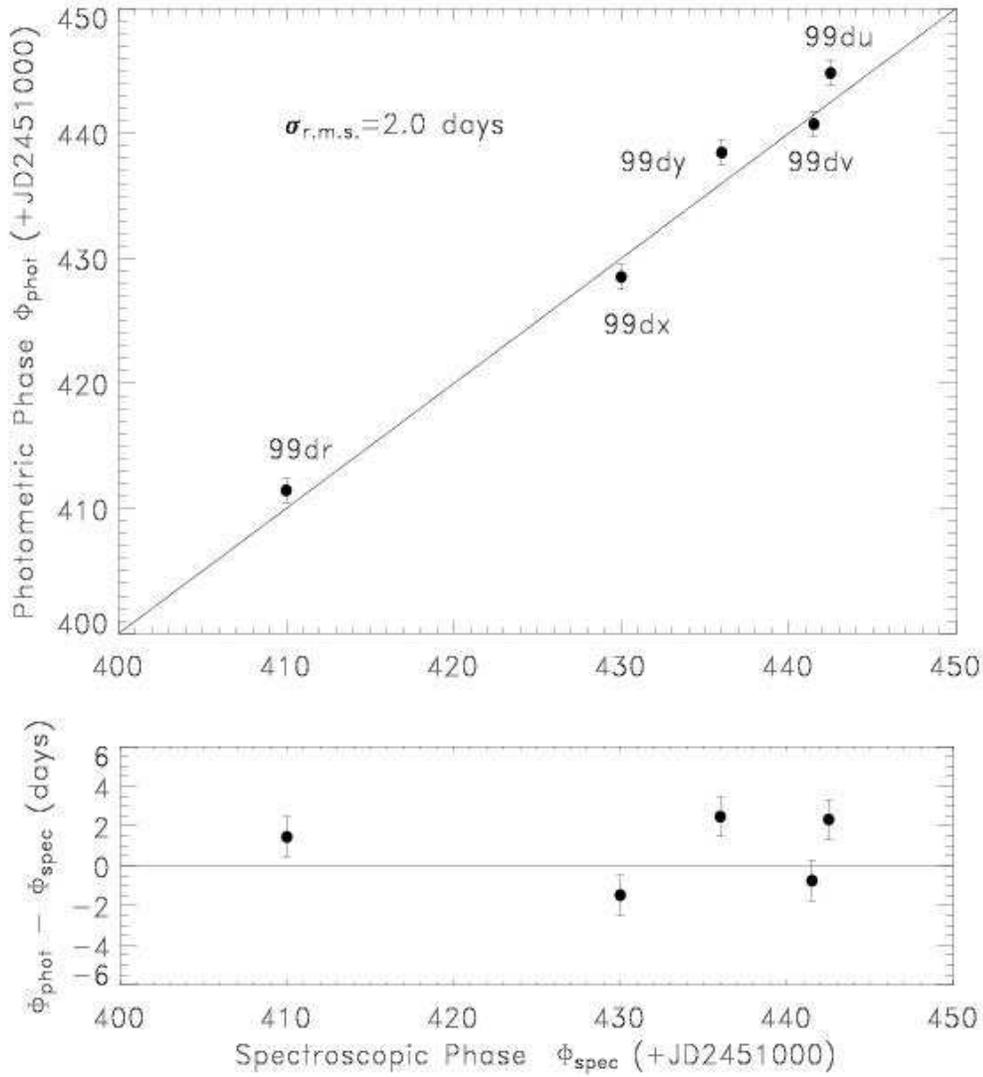}}
\caption{Photometric phase relative to spectroscopic phase for the
  five photometrically followed-up \Ia of our sample (top panel), and
  residuals (bottom panel).  Errors shown are for the photometric
  phase only and assume a conservative $\pm 1$ day error. The
  photometric phase is from fit of the restframe B-band light-curve
  and the spectroscopic phase from $\chi^2$ minimization fitting of
  observed spectra with \Ia spectral templates}
\label{fig19}
\end{figure*}

We have argued that the fitting procedure used in this paper allows us
to constrain the phase within typically $\pm 3$ days. We can correlate
this 'spectroscopic' phase with the photometric phase obtained from
B-band photometry for five out of the twelve SNe of our sample which
have been photometrically followed-up (SN~1999dr, SN~1999du,
SN~1999dv, SN~1999dx and SN~1999dy). The photometric phases are
derived by constructing and fitting the light-curves for these SNe, as
described in our forthcoming paper.  Preliminary results are given in
Table \ref{tab6}. A $\pm$ 1 day error bar on the photometric phase is
assigned to each supernova. This value is likely to overestimate the
true error as light-curve fitting constrains the date of maximum
within a fraction of a day. It however accounts for possible
systematics in the photometric analysis.  We find good agreement
between the photometric and spectroscopic dates of maximum for the
subset. When a correction for time dilation is applied, the agreement
is even better, except for SN~1999dr. As already mentioned, the
spectral phase determination of this \Ia well after maximum light is
likely to be less precise than for other supernovae.  Figure
\ref{fig19} shows the photometric phase (column 2 of Table \ref{tab6})
as a function of the spectroscopic one (column 3). The rms dispersion
is $\sigma_{rms}=$2.0 days. This is consistent with the $\pm 3$ days
range we adopted for the spectroscopic phase. This value reduces to
$\sigma_{rms}=1.7$ day if times dilation corrected values are
considered (column 5, excluding the uncertain SN~1999dr).

We also compare the phases obtained by fitting peculiar templates
(Table \ref{tab7}) with the photometric phase for the 5 followed-up
September supernovae. The dispersion increases from 2.0 days
('normal') to $\approx$ 6.5 days (peculiar). Again, this tends to show
that these supernovae are 'normal'.

\subsection{Host results}

\begin{figure*}
  \resizebox{14cm}{!}{\includegraphics{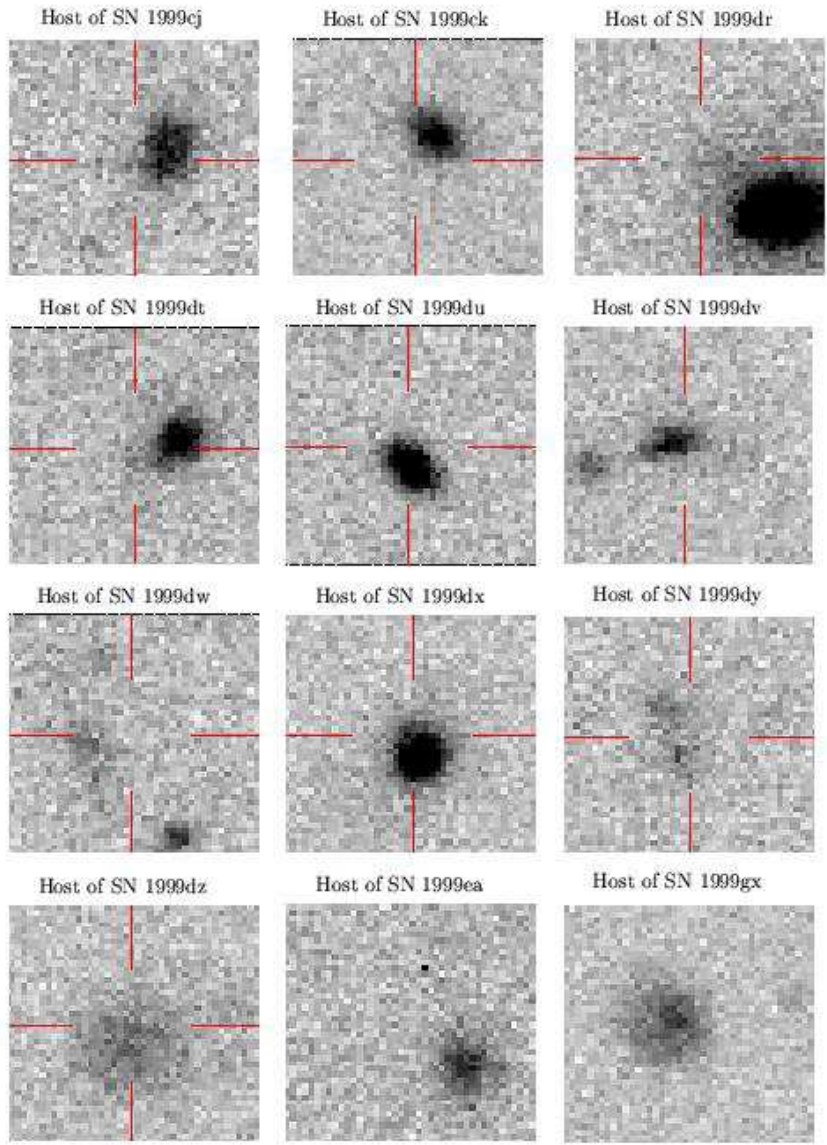} }
\caption{WFS reference images of host galaxies. Each vignette is a 0.25$\times $0.25 square-arcmin cutout of the full CCD, centered  on the position where the SN exploded (indicated by a cross). All images are in g', except for SN~1999cj (B image) and SN~1999ck (r' image).}
\label{fig20}
\end{figure*}

Figure \ref{fig20} shows WFS reference images of host galaxies of the
twelve supernovae presented in this paper. Each vignette is a
0.25$\times $0.25 square-arcmin g' image, except for SN 1999cj (B-band
image) and SN 1999ck (r' image). Pixel size is 0.33 arsec.  The cross
indicates the position where the SN has exploded.

Table \ref{tab5} summarizes the host identification for each candidate
(cf. Sect. 4.3). The features used for the identification are presented in
columns 2 and 3. Specific comments for each galaxy are given in column
4 while in column 5, our best determination for the host type is
given. In column 6 we indicate the confidence we have in the
identification in the form of an index $n_c$ whose values can be 0
(type unsure), 0.5 (average confidence), or 1 (high confidence). Low
confidence in the identification results from a poor signal-to-noise
ratio, a fact that usually correlates with a 'high' redshift and a
poor sky subtraction, or with a too low contamination of the supernova
by the host.  Finally, column 7 gives the 'best-fit type' given by
${\cal SN}$-fit (see \S 4.3). It is clear that in most cases, this
'best-fit type' agrees with the direct identification.

For the sake of completeness, we have also computed u'-g', B-V and
g'-r' colors for the twelve host galaxies. u', g' and r' magnitudes
have been taken from the Sloan Digital Sky Survey (SDSS) on-line
catalog.  B-V colors have been computed using color equations given by
\citet{Fukugita96}.  E(B-V) has been computed for each galaxy from the
r' Galactic extinction given in the SDSS catalog and an extinction
correction has been applied to all colors.  Results are shown in Table
\ref{tab8}.

These derived color values can then be compared to the galaxy colors
computed by \citet{Frei94} for their four Hubble galaxy types (E, Sbc,
Scd and Im). To connect this classification to the one we have
adopted, we associate their type E to our type 0, Sbc to our type 1
and Scd and Im to our type 2. We assume that a Sa galaxy (type 1 in
our classification) has colors falling in between their E and Sbc
types.  \citet{Frei94} computed galaxy colors using the galaxy energy
distributions compiled for each Hubble type by \citet{Coleman80}.
Color values are given at four different redshifts ($z=0, 0.1, 0.4$
and 0.6). For the purpose of comparison, we have extrapolated Frei and
Gunn's colors at the redshifts of our hosts, for each Hubble type.

Based on this color comparison, we confirm the identification made
from spectra for eight hosts (SN~1999cj, SN~1999ck, SN~1999dr,
SN~1999du, SN~1999dv, SN~1999dx, SN~1999dy and SN~1999dz) out of
twelve. Two hosts (SN~1999ea and SN~1999gx) have colors marginally
consistent with the type derived from their spectra (at least one color
does not agree with the spectral type). The colors of SN 1999dt and SN
1999dw hosts are consistent with type 0 (two colors out of three hint
toward a 0-type host) while they have been identified as type 1 from
their spectra.

\section{Discussion}
 
Based on the classification results given in Table \ref{tab5}, two out
of twelve hosts are identified as type 0, seven as type 1 and three as
type 2, corresponding to a fraction of 17\%, 58\% and 25\% of the
sample respectively.  Although we are limited by the small statistics
of our sample, we can compare our results with the host statistics of
\citet{Sullivan03} on two samples of low and high redshift. The
low-$z$ sample is based on the Calan-Tololo sample \citep{Hamuy96b}
and on the \citet{Riess99a} local sample. All supernovae in this
sample have $z<0.01$. The high-redshift sample is based on the 42 SNe
of the SCP \citep{Perlmutter99} with redshift ranging from $z=0.18$ to
$z=0.83$.  Our fractions are roughly consistent with the statistics of
hosts in the low-$z$ sample of Sullivan \etal, who find 12\%(0),
56\%(1) and 32\%(2). In their high-$z$ sample, the fraction of E/S0s
levels up with early-type spirals whereas the late-type spiral
fraction increases: 28\%(0), 26\%(1) and 46\%(2). This trend is not
seen in our sample.  To check if consistency can be found on a subset
of the high-$z$ sample with similar average redshift as ours, we have
selected the nine hosts with redshift ranging from $z=0.172$ and
$z=0.43$ with an average redshift of $<z>\approx 0.34$. However, the
same trend as for the total high-$z$ sample is clearly seen, with even
more type 0.  As the number of \Ia of our sample is small, possible
misidentification of one host galaxy could significantly change our
numbers. If, say, we have classified as type 1 an actually type 0
galaxy (SN~1999dr for example, which has a red u'-g' color value
consistent with the one for type 0), we would get 25\%(0), 50\%(1) and
25\%(2). However, this does not influence the fraction of late-type
spirals and we would need to have misclassified one type 2 as, for
instance, a type 1 to alleviate the discrepancy.  This is rather
unlikely as strong emission lines seen in type 2 spectra should be a
clear signature. In our sample, these strong emissions only appear for
SN~1999du, SN~1999dv and SN~1999dy.  Clearly, complementary
information as galaxy colors and a larger set of intermediate redshift
\Ia would help to conclude.

Taking into account various biases intrinsic to supernova searches
that may affect the peculiarity rate of SN Ia, \citet{Li01b} find that
36$\pm$9\% of nearby \Ia are peculiar.  Contrasting with this result,
no peculiar \Ia have been found in the high-$z$ samples of
\citet{Riess98b} and \citet{Perlmutter99}. This might indicate
evolution with redshift in the \Ia populations, even though
identification of peculiar objects at high redshift is rendered
difficult due to insufficient signal-to-noise and potentially severe
age-bias: spectral differences between 'normal' and over-luminous \Ia
tend to vanish after maximum light, which introduces an 'age-bias' in
the detection of such over-luminous \Ia \citep{Li01a,Li01b}.  Assuming
the fraction expected in nearby surveys holds at intermediate
redshift, we would expect between 3 and 5 peculiar supernovae in our
sample. Our analysis yields $0\pm 1$ peculiar \Ia similarly to the
high-$z$ sample.  However, magnitude-limited surveys such as ours
underestimate the rate of under-luminous \Ia (SN~1991bg-like objects)
as opposed to local surveys which are rather distance-limited.
Over-luminous (SN~1991T-like) objects are usually seen in dusty
star-forming regions \citep{Li01a}. Although $\sim $0.4 mag brighter than 'normal'
\Ia, they suffer from heavy extinction and might be under-represented
in our sample.  Moreover, their detection might be affected by the
age-bias.  If we assume one of our \Ia has been incorrectly classified
as a 'normal' (for example, SN~1999cj, SN~1999dr whose spectra have
been taken well after maximum, or possibly SN~1999gx), the peculiarity
rate in our sample would be $~\sim 8-9$ \%, a value consistent with
the 12 \% expected for a magnitude-limited survey with a 20 day
baseline, with an extra R-band extinction for SN~1991T-like objects of
0.8 mag and an age-bias cutoff of 7 days \citep{Li01b}.

Following the first results on cosmological parameters based on the
observation of distant supernovae, systematic searches for \Ia have
been undertaken on various telescopes and several samples exist today.
Although aiming at observing higher redshift \Ia than ours, they often
contain a few \Ia in the intermediate redshift range and fairly good
signal-to-noise spectra have been obtained.  \citet{Riess98b} early
sample includes five \Ia in the range $0.15<z<0.3$, while
\citet{Perlmutter99} have two \Ia around 0.17. More recently
\citet{Barris04} present spectroscopy for 23 supernovae discovered
during the IfA Deep Survey, among which none are in the range
$0.15<z<0.3$. \citet{Lidman05} find two \Ia in this redshift range.
The large scale program ESSENCE \citep{Matheson05} presents nine
spectra of \Ia between $z=$0.15 and $z=$0.3 after two years of
systematic searches. Even more recently, discovery of supernovae in
the SDSS survey has been reported, among which seven have been
identified as \Ia in the same redshift range \citep{sako05}. Including
our spectra, the number of \Ia spectroscopically observed between 0.15
and 0.3 amounts to thirty, all with similar signal-to-noise.

Identification is always based on visual detection of characteristic
spectral features such as SiII 6355\AA$\ $ and/or comparison to well
observed local \Ia. This comparison is direct \citep{Riess98b} or
involves a template fitting procedure similar as ours
\citep{Barris04,Matheson05}.  In \citet{Riess98b}, differences between
templates and real spectrum are small and are comparable to what we
obtain with our set.  The fitting result is of poorer quality in
\citet{Barris04} and \citet{Matheson05}, however good enough for the
sole purpose of identification of \Ia features. No galaxy subtraction
has been attempted for the ESSENCE spectra, which degrades the visual
aspect of the result when the full spectrum is fitted by local \Ia
templates. Note that one out of the nine spectra of the
intermediate-redshift subsample of \citet{Matheson05} shows strong
similarities with SN~1991T, whereas all the five \Ia of the
\citet{Riess98b} sample are best compared to 'normal' \Ia templates.
These results are consistent with our finding of none, or possibly
one, peculiar \Ia in our intermediate redshift sample.

\section{Conclusion}

We have presented twelve spectra of \Ia supernovae taken at the
William Herschel Telescope and the Nordic Optical Telescope in April,
September and October 1999 during a search for \Ia at intermediate
redshifts. Five of these SN have redshifts between $0.15<z<0.3$. This
set provides high signal-to-noise spectra in a still largely
unexplored redshift range.

The identification was based on a $\chi^2$-minimization using a
database of galaxy and \Ia templates, which allow us to model the
observed spectra in a consistent way.  Determination of the
spectroscopic phase is reliable.  Comparison with the photometric
phase derived from the light-curves of five \Ia that have been
photometrically followed-up gives an estimated $\pm 2$ day
uncertainty when measuring the phase from the matching to the
spectroscopic database.

Spectral analysis shows that most of the objects found during this
campaign are clearly spectroscopically 'normal'. The observed spectra
are best fitted with templates of local 'normal' SN Ia, even if the
possibility remains that SN~1999gx is a peculiar over-luminous
supernova. In that case, it would be the most distant peculiar
supernova observed so far. Velocity measurements of characteristic
absorption lines such as CaII, SII and SiII are consistent with the
same measurements on 'normal' \Ia, with SN~1999dr having a slightly
higher value than average.

The peculiarity rate in our sample is inconsistent with the rates
predicted in local surveys but similar to what is observed in the
high-$z$ sample of \citet{Riess98b} and \citet{Perlmutter99}.  But,
although unlikely, possible misidentification of a SN~1991T-like
object as a 'normal' \Ia in our sample would reconcile observations
with the predicted local rate, provided that those objects suffer
extra extinction.

We conclude that the physical properties of the intermediate redshift
\Ia presented in this paper are very similar, as far as the
spectroscopic analysis is concerned, to the properties of their low
and high redshift counterparts.

\begin{acknowledgements}
  The observations described in this paper were primarily obtained as
  visiting/guest astronomers at the INT and WHT, operated by the Royal
  Greenwich Observatory at the Spanish Observatorio del Roque de los
  Muchachos of the Instituto de Astrofísica de Canarias, and the
  Nordic Optical 2.5 m telescope. We thank the dedicated staffs of
  these observatories for their assistance in pursuit of this project.
  We also acknowledge G. Altavilla for useful comments on the
  manuscript and G. Garavini for his help in producing Fig. 18. We
  thank S. Nobili for providing us with her set of spectral templates.
  Some of the spectral templates used in ${\cal SN}$-fit database have
  been kindly provided to us by T. Matheson and A. Filippenko.
     
\end{acknowledgements}
\bibliographystyle{aa}
\bibliography{bibi}

\newpage

\vfill
\eject

\clearpage
\newpage

\newpage

\newpage

\newpage

\newpage

\newpage

\newpage

\clearpage

\clearpage

\clearpage

\clearpage

\clearpage

\newpage

\small
\begin{longtable}[l]{cccccccccc}
\caption[c]{\label{tab1} Log of the Spectroscopic Observations}\\
\hline \hline
SN & RA & Dec. & UT Date & [R] or g' & UT Date of  & Telescope/ & Grating/ & Dispersion & Slit (arcsec)/\\
& (2000) & (2000) & of Discovery  & mag$\dagger$ & Spectroscopy & Instrument & Grism & (\AA /pix) & Par. angle$\ddagger$\\
\hline \endfirsthead
1999cj & 10 14 02.46 & - 0 14 34.0$^\ast$ & Apr 17 & [19.2] & Apr 19 & WHT/ISIS & R158B,R158R & 1.6/2.9 & 1.0/no\\
1999ck & 14 08 56.59 & - 0 05 50.7 & Apr 16 & [21.9] & Apr 19 & WHT/ISIS & R158B,R158R & 1.6/2.9 & 1.0/yes\\
1999dr & 23 00 17.56 & - 0 05 12.5 & Sep 1 & 22.1 & Sep 12 & WHT/ISIS &R158R  & 2.9 & 1.5/no\\
1999dt & 00 45 42.29 & + 0 03 22.2 & Sep 4 & 23.5 & Sep 12 & WHT/ISIS &R158R  & 2.9 & 1.5/no\\
1999du & 01 07 05.94 & - 0 07 53.8 & Sep 8 & 22.8 & Sep 12 & WHT/ISIS &R158R  & 2.9 & 1.5/yes\\
1999du & `` & `` & `` & `` & Oct 2 & NOT/ALFOSC & \#4 & 3.0 & 2.5/yes\\
1999dv & 01 08 58.96 & + 0 00 24.8 & Sep 9 & 21.8 & Sep 13 & WHT/ISIS &R158R  & 2.9 & 1.5/yes \\
1999dv & `` & `` & `` & `` & Oct 2 & NOT/ALFOSC & \#4 & 3.0 & 1.0/yes\\
1999dw & 01 22 52.80 & - 0 16 20.8 & Sep 7 & 24.1 & Sep 12 & WHT/ISIS &R158R  & 2.9 & 1.5/no \\
1999dx & 01 33 59.45 & + 0 04 15.3 & Sep 8 & 22.2 & Sep 14 & WHT/ISIS &R158R  & 2.9 & 1.5/no \\
1999dy & 01 35 49.53 & + 0 08 38.3 & Sep 8 & 21.7 & Sep 14 & WHT/ISIS &R158R  & 2.9 & 1.5/no \\
1999dz & 01 37 03.24 & + 0 01 57.9 & Sep 8 & 23.4 & Sep 13 \& 14  & WHT/ISIS &R158R  & 2.9 & 1.5/yes\\
1999ea & 01 47 26.09 & - 0 02 07.2 & Sep 9 & 23.3 & Sep 13 & WHT/ISIS &R158R  & 2.9 & 1.5/no \\
1999gx & 00 34 15.47 & + 00 04 26.2 & Sep 6 & 23.3 & Sep 13 & WHT/ISIS &R158R  & 2.9 & 1.5/yes \\
\hline
\end{longtable}

\footnotesize{Notes:}\\
\hspace*{5mm}\footnotesize{$^\dagger$ Approximate magnitude at discovery}\\
\hspace*{5mm}\footnotesize{$^\ddagger$ Parallactic Angle}\\
\hspace*{5mm}\footnotesize{$^\ast$ Note that, due to a typographic
mistake, the dec. for SN~1999cj differs by 31.5'' from the value given in \citet{Hardina}}\\

\small
\begin{longtable}[c]{ccccccccc}
\caption[c]{\label{tab2} Results from ${\cal SN}$-fit}\\
\hline \hline
SN     & Telescope & Best fit SN& \% (host)$^\dagger$ & $z_f$  & $z_h$  & Spectroscopic & Reduced & d.o.f.$^\ast$ \\
name & & template (reference) & & & & phase$^\ddagger$ & $\chi^2$ &\\ 
\hline
1999cj &   WHT   & 92A (Kirshner \etal 1993)& 31(Bulge)           & 0.355       & 0.362  & +9   & 1.00 & 859\\
1999ck &   WHT   & 94D (Patat \etal 1996) & 48(Sa)                & 0.434       & 0.432  & -9   & 1.17 & 502 \\
1999dr$^1$ &  WHT &94D (Patat \etal 1996) &  48(own galaxy)       & 0.174       & 0.178  & +24  & 0.95 & 935\\
1999dt & WHT &94D (Patat \etal 1996) &  80(Sb)                & 0.438       & 0.437  & -9   & 1.24 & 623 \\
1999du$^2$ &  WHT &99ee (Hamuy \etal 2002) &  65(Stb4)            & 0.258       & 0.260  & -9  & 0.87 & 853\\
1999dv &  WHT &03du (Anupama \etal 2005) &  11(Stb1)              & 0.177       & 0.186  & -7   & 0.99 & 936\\
1999dw &  WHT &99ee (Hamuy \etal 2002) &  63(Sa)                  & 0.464       & 0.460  & -4   & 0.75 & 705\\
1999dx &  WHT &92A (Kirshner \etal 1993) &  77(S0)                & 0.270       & 0.269  & +5   & 1.66 & 935 \\
1999dy &  WHT &96X (Salvo \etal 2001) &  23(Sb)                   & 0.202       & 0.215  &  0   & 1.14 & 934\\
1999dz &  WHT &99ee (Hamuy \etal 2002) &  79(Sb)                  & 0.480      & 0.486  & -4   & 1.17 & 671\\
1999ea &  WHT &94D (Patat \etal 1996) &  42(Sa)                   & 0.398       & 0.397  & -8   & 1.12 & 697\\
1999gx      & WHT & 94D (Patat \etal 1996) & 64(Sa)                    & 0.487       & 0.493 & -4 & 0.93 & 871\\
1999du$^1$ &  NOT &92A (Kirshner \etal 1993) & 28(own galaxy)         & 0.250       & 0.260 & +6 & 0.81 & 572\\
1999dv  &  NOT &92A (Kirshner \etal 1993) & 6(Sc)                 & 0.181 & 0.186 & +9 & 1.15 & 637\\
\hline
\end{longtable}

\footnotesize{Notes:}\\
\hspace*{5mm}\footnotesize{$^\dagger$ \%(host) = $\beta/(\alpha + \beta)\times 100$ (see text)}\\
\hspace*{5mm}\footnotesize{$^\ddagger$ Typical error is $\pm 3$ days}\\
\hspace*{5mm}\footnotesize{$^\ast$ Number of degrees of freedom}\\
\hspace*{5mm}\footnotesize{$^1$ The observed galaxy has been used
  for the fit}\\
\hspace*{5mm}\footnotesize{$^2$ Best solution when emission lines are
  ignored: 99ee -9 days, 39\% (Sb), $z_f$=0.261, reduced $\chi^2=0.78$, d.o.f.=936}\\

\small

\begin{longtable}[c]{ccccc}
\caption{\label{tab4} SN line velocities}\\

\hline
\hline
SN & Telescope & $v_{\mathrm{CaII}}^a$ & $v_{\mathrm{SII}}^b$ & $v_{\mathrm{SiII}}^c$ \\ \hline
1999cj 
& WHT & -12400 & &   \\
1999ck
& WHT  &-14700 & &   \\
1999dr 
& WHT&  &  & -10350 \\
1999dt 
& WHT&  &  &   \\
1999du 
& WHT&  &  &   \\
1999dv 
& WHT&  & -11200 & -11300  \\
1999dw 
& WHT& -17400 &  &   \\
1999dx 
& WHT& -15650  & -9850   & \\
1999dy 
& WHT&  & -11500 &  \\
1999dz 
& WHT& -16900 &  & \\
1999ea 
& WHT& -15750 &  &  \\
1999gx & WHT & -16200 & & \\
1999du & NOT & -15200 & &  \\

\hline
\end{longtable}

\footnotesize{Notes:}\\
\hspace*{5mm}\footnotesize{$^a$ Ca H\& K $\lambda 3945$ in km/s;
typical error is 500 km/s}\\
\hspace*{5mm}\footnotesize{$^b$ SII $\lambda 5640$ in km/s; typical error is 200 km/s}\\
\hspace*{5mm}\footnotesize{$^c$ SiII$\lambda 6355$ in km/s; typical error is 200 km/s}\\

\small
\begin{longtable}[c]{cccccccccc}
\caption[]{\label{tab7} Results from${\cal SN}$-fit using peculiar \Ia templates}\\
\hline\hline SN & Telescope & \multicolumn{4}{c}{Peculiar} &
\multicolumn{3}{c}{Normal} &
F-test\\
& & \multicolumn{4}{c}{------------------------------------------} & \multicolumn{3}{c}{-----------------------------} &probability (\%) \\
& &  Best-fit & Phase & $\chi^2$ & d.o.f. & Phase & $\chi^2$ & d.o.f. & \\
\hline
1999cj & WHT & 99aa$^1$ & +14 & 1.025 & 859 & +9 & 1.005 & 859 & 57\\
1999ck & WHT & 99aa & -7  & 1.309 & 533 & -9 & 1.171 & 502 & 1.9\\
1999dr & WHT & 00cx$^2$ & +30 & 1.100 & 900 & +24 & 0.949 & 935 & $2.2\ 10^{-3}$\\
1999dt & WHT & 99aa & -7  & 1.523 & 667 & -9 & 1.245 & 623 & $6\ 10^{-4}$\\
1999du & WHT & 99aa & -7  & 0.945 & 894 & -9 & 0.867 & 853 & 1.7\\
1999dv & WHT & 99aa & -7  & 1.443 & 833 & -7  & 0.994 & 436 & $2\ 10^{-20}$\\
1999dw & WHT & 99aa & -7  & 0.764 & 643 & -4 & 0.746 & 705 & 52\\
1999dx & WHT & 99aa & -3  & 1.825 & 935 & +5 & 1.654 & 935 & 0.4\\
1999dy & WHT & 99aa & -3  & 2.508 & 932 &  0 & 1.136 & 934 & $<10^{-20}$\\
1999dz & WHT & 00cx & -3  & 1.263 & 563 & -4 & 1.169 & 971 & 0.8\\
1999ea & WHT & 99aa & -11 & 1.321 & 804 & -8 & 1.117 & 697 & $10^{-2}$\\
1999gx & WHT & 91T$^3$ & -12  & 0.939 & 674 & -4 & 0.938 & 794 & 97\\
1999du & NOT & 00cx & +14 & 0.883 & 336 & +6 & 0.812 & 572 & 2.7\\
1999dv & NOT & 99aa & +14 & 1.429 & 542 & +9 & 1.153 & 637 & 4$\ 10^{-5}$\\
\hline
\end{longtable}

\footnotesize{Notes:}\\
\hspace*{5mm}\footnotesize{$^1$ Spectral template from Garavini et al. (2004)}\\
\hspace*{5mm}\footnotesize{$^2$ Spectral template from Li et al (2001c)}\\
\hspace*{5mm}\footnotesize{$^3$ Spectral template from Mazzali et. al. (1995)}\\

\small
\begin{longtable}[c]{cccccc}
\caption{\label{tab6} Spectral and light-curve dates of maximum for the five photometrically followed-up \Ia}\\
\hline\hline
SN & HJD(civil date)$\dagger$ & \multicolumn{2}{c}{Uncorrected maximum date$\ddagger$} & \multicolumn{2}{c}{Corrected maximum date$\ast$}\\
&  &\multicolumn{2}{c}{---------------------------------------} & \multicolumn{2}{c}{---------------------------------------}\\
&  & WHT & NOT & WHT & NOT\\\hline

1999dr & 2451411.45 (20/8) & 19/8 & ... & 14/8 & ...\\
1999du & 2451444.83 (23/9) & 21/9 & 26/9 & 23/9 & 24/9\\
1999dv & 2451440.75 (19/9) & 20/9 & 23/9 & 21/9 & 21/10\\
1999dx & 2451428.52 (07/9) & 09/9 & ... & 08/9 & ...\\
1999dy & 2451438.47 (16/9) & 14/9 & ... & 14/9 & ...\\
\hline
\end{longtable}

\footnotesize{Notes:}\\
\hspace*{5mm} {\footnotesize $\dagger$ Date of B-band light-curve maximum}\\
\hspace*{5mm}{\footnotesize $\ddagger$ Date estimated from the date of
  spectral observation corrected for the best-fit spectral phase}\\
\hspace*{5mm}{\footnotesize $\ast$ Date estimated from the date of
  spectral observation corrected for the best-fit spectral phase. Correction for time
dilation effects has been applied}

\small

\begin{longtable}[c]{ccccccc}
\caption{\label{tab5}SN host spectral classification.}\\
\hline
\hline
SN &  T3850$^\dagger$ & Other  &  Comments & Host Spectral & Confidence 
 & ${\cal SN}$-fit  \\ 
 &  & indicators$^\ast$ &  & type$^\ddagger$& index $n_c^\star$ &  \\
\hline
1999cj   & - & [OII], s. Mgb                            & red, likely E/S0 & 0      & 0.5 & Bulge   \\
1999ck   & - & s. [OII], w. H$\beta$, w. [OIII]         & poor gal. signal & 1      & 0.5 & Sa      \\
1999dr   & n. & s. H, K; B4000; s. Mgb                &  no emission       & 1      & 0.5 & Own gal.      \\
1999dt   & n. & s. [OII]; s. H, K; B4000; s. H$\gamma$  & Sa/Sb            & 1      & 1   & Sb      \\
1999du   & - & s. [OII], H$\beta$, [OIII]               & clear ident. Sc  & 2      & 1   & Stb     \\
1999dv   &- & s. H$\beta$, [OIII]                     & poor gal. signal   & 2      & 0.5 & Stb      \\
1999dw   & - & w. [OII]                                & poor S/N          & 1      & 0   & Sa      \\
1999dx   & y.& s. H, K; B4000; w. [OII]; w H$\gamma$    & clear ident. E/S0& 0      & 1   & S0      \\
1999dy   & - & [OII]; [OIII]                            & faint            & 2      & 0.5 & Sb      \\
1999dz   & n. &  w. [OII]; H, K                         & likely Sa/Sb     & 1      & 0.5 & Sb      \\
1999ea   & n. & s. [OII]; [OIII]; H$\beta$; H, K       & Sa/Sb             & 1      & 1   & Sa      \\
1999gx   & -  & [OII]; H, K                             & Sa/Sb            & 1      & 0.5 & Sa      \\
\hline
\end{longtable}

\footnotesize{Notes:}\\
\hspace*{5mm}\footnotesize{$^\dagger$ T3850=3850\AA $\ $ CN trough present: y.=yes, n.= no; - = not identified}\\
\hspace*{5mm}\footnotesize{$^\ddagger$ Type 0: spheroidal (E/S0); Type 1: early-type spiral (Sa/Sb); Type 2: late-type spiral (Sc/Starburst)}\\
\hspace*{5mm}\footnotesize{$^\ast$ B4000=4000\AA $\ $ break, s.=strong, w.=weak}\\
\hspace*{5mm}\footnotesize{$^\star$ High confidence: $n_c$=1; Average confidence: $n_c$=0.5; Low confidence: $n_c$=0}\\

\small
\begin{longtable}[c]{ccccc}
\caption{\label{tab8} Extinction Corrected Host Galaxy SDSS Colors}\\
\hline\hline
SN & E(B-V) & B-V & g'-r' & u'-g' \\
\hline
1999cj & 0.035 & 1.53 & 1.38 & 2.12\\ 
1999ck & 0.04  & 1.33 & 1.16 & 0.74\\
1999dr & 0.06  & 1.31 & 1.15 & 2.12\\
1999dt & 0.015 & 1.53 & 1.38 & 3.05\\
1999du & 0.03  & 0.72 & 0.53 & 0.45\\
1999dv & 0.035 & 0.67 & 0.47 & 1.02\\
1999dw & 0.04  & 1.67 & 1.53 & 3.05\\
1999dx & 0.04  & 1.59 & 1.44 & 2.69\\
1999dy & 0.03  & 0.88 & 0.70 & 0.25\\
1999dz & 0.04  & 1.62 & 1.47 & 1.39\\
1999ea & 0.035 & 1.03 & 0.85 & 1.22\\
1999gx      & 0.02  & 1.18 & 1.01 & 0.86\\
\hline
\end{longtable}

\newpage

\appendix
\section{${\cal SN}$-fit spectral database}

A total of 242 spectral templates have been collected from various
sources into the ${\cal SN}$-fit database used for the analysis of the
spectra presented in this paper. These include 167 type Ia spectra
(both 'normal' and peculiar\footnote{Peculiar Ia include 91bg-like and
  91T-like \Ia}), 64 type Ib/c and II spectra and 11 galaxy spectra.
Most of the type Ia and type II templates originate from the Asiago
Catalog\footnote{\tt http://web.pd.astro.it/supern/} and have been
retrieved from the SUSPECT\footnote{\tt
  http://bruford.nhn.ou.edu/$\sim$suspect/} database
\citep{Richardson02} with the exception of SN~1981B, SN~1992A, as well
as some spectra of SN~1990N and SN~1991T. Most type Ib/c templates are
from Matheson (private communication and \citet{Matheson01}). Galaxy
templates are taken from \citet{Kinney96} and cover a large spectral
range (from 1200 to 10000 \AA).  Morphological types include bulge, E,
S0, Sa, Sb, Sc, and starbursts (Stb) with various amount of reddening
\citep{Calzetti94,Kinney96}.
  
The selected supernova spectra have: i) a large wavelength coverage in
the visible - typically 4000 to 8000 \AA$\ $ restframe, ii) a good
signal to noise, iii) a continuous energy distribution in the
wavelength range considered (templates with gaps in their energy
distribution have been discarded). All spectra have been de-redshifted
and de-reddened when necessary. The spectra phases cover the range
-15,45 days - with respect to maximum light - as uniformly as
possible. A few later phase spectra (phase greater than 45 days) have
been included to the database for peculiar \Ia (4 spectra) and
Ib/Ic/II (8 spectra). Table \ref{tab9} and \ref{tab10} show the
supernova templates available in ${\cal SN}$-fit database along with
the corresponding number of spectra, for type Ia and Ib/Ic/II
respectively.
  
\begin{figure*}
 \resizebox{14cm}{!}{\includegraphics{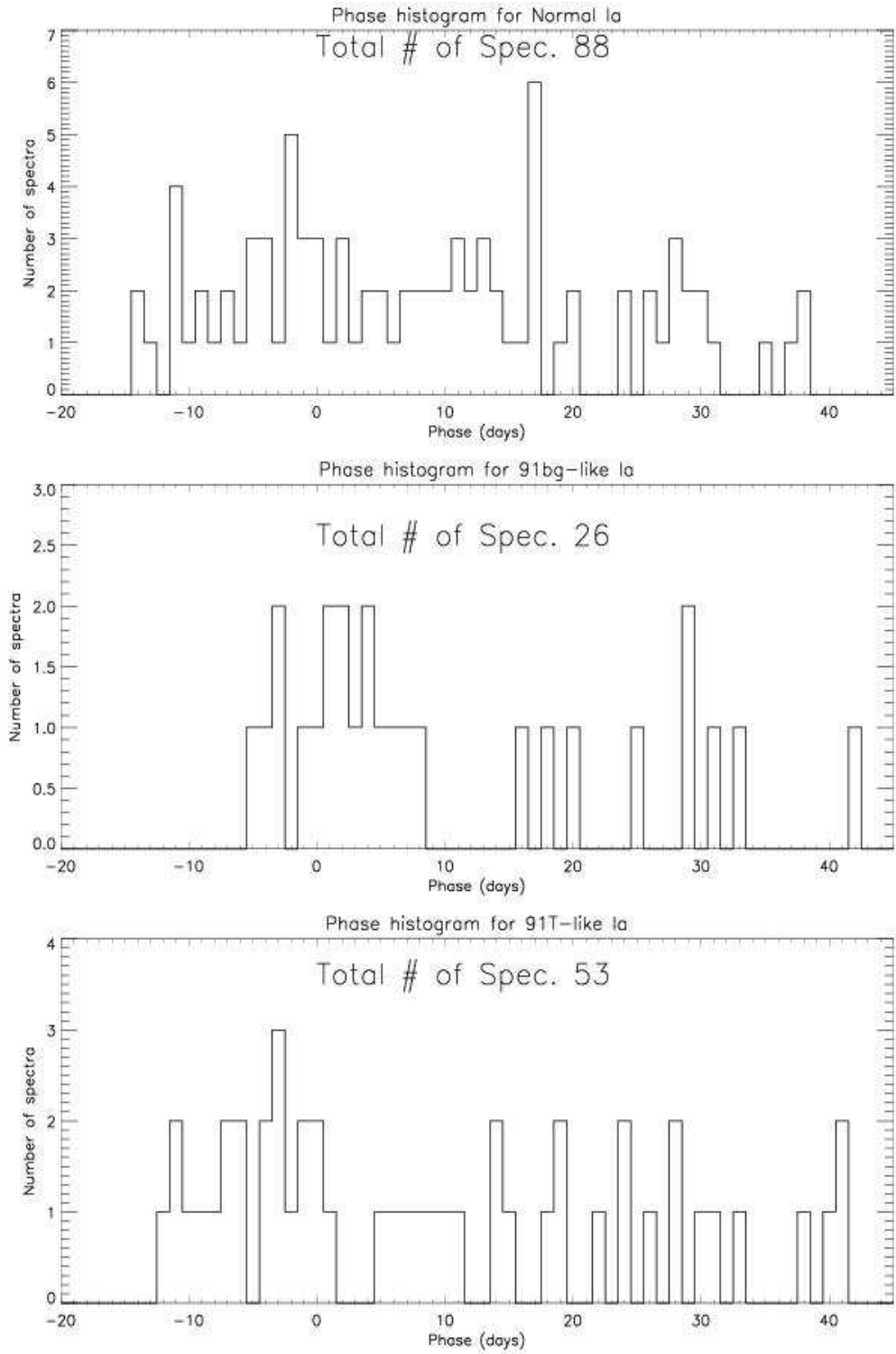} }
\caption{Phase distribution of type Ia templates in ${\cal SN}$-fit database. Up: Normal \Ia. Middle: 91bg-like \Ia. Bottom: 91T-like \Ia.}
\label{histoIa}
\end{figure*}

\begin{figure*}
 \resizebox{14cm}{!}{\includegraphics{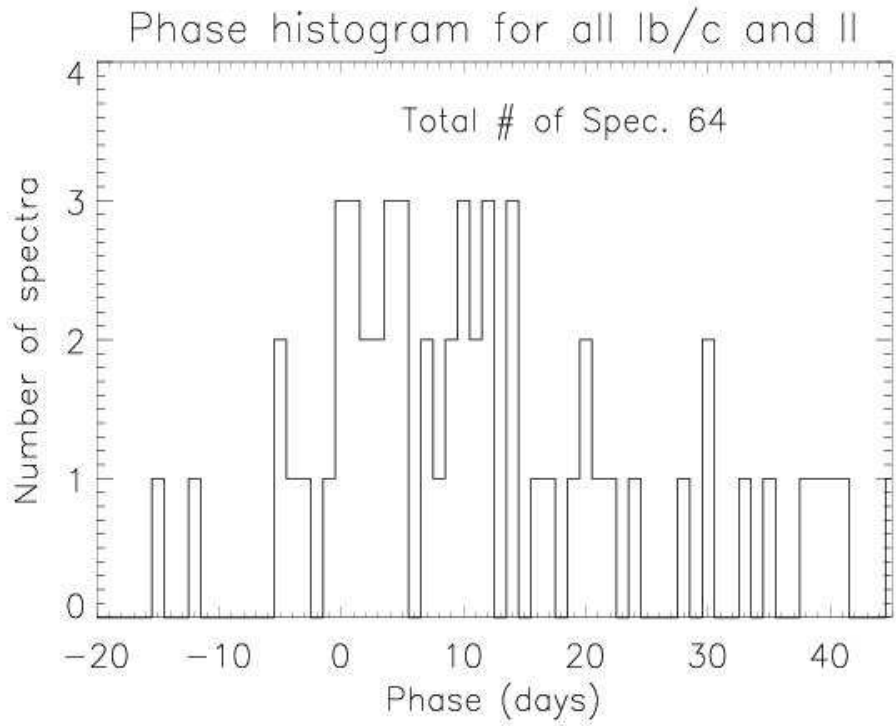} }
\caption{Phase distribution of type Ib, Ic and II templates in ${\cal SN}$-fit database.}
\label{histoIbcII}
\end{figure*}

The phase coverage of \Ia and SN~Ib/Ic/II templates available in the
database is shown as histograms in Fig. \ref{histoIa} and
\ref{histoIbcII}. The phase ranges from -15 to +45 days. Note however
that the total number of spectra indicated in the figures takes into
account the spectral templates with phases greater than 45 days, which
are not shown here.
  
In Fig.  \ref{histoIa}, both the phase distribution for 'normal' \Ia
(top panel) and peculiar \Ia (91bg-like \Ia: middle panel; 91T-like
\Ia: bottom panel) are shown. Phases for normal \Ia templates are
uniformly spread between -15 and 40 days, except for a gap of three
days after day 20. This gap will be filled in the future as more
spectra of low-redshift supernovae become available.  Note also the
absence of spectra at day -12.

91bg-like \Ia templates are found mostly around maximum. The sampling
is smooth between -5 and +8 days, with the exception of day -2 for
which templates are missing. No template is available between day 9
and day 15. The distribution of 91T-like \Ia templates peaks a few
days before maximum. Note the absence of spectra at days 2, 3 and 4.
After day +11, the sampling is scarcer with a phase resolution of 2
days up to about day 30.
  
For completeness, we also show in Fig. \ref{histoIbcII} the phase
distribution for type Ib/Ic and II. It can be seen that spectral
phases for these types are best represented in the database between -5
and + 15 days.
  
Note that ${\cal SN}$-fit database is in constant evolution. New
spectra are regularly included as they become available to the
community.

\small
\begin{longtable}[c]{cccc}
\caption{\label{tab9} SN~Ia template spectra in ${\cal SN}$-Fit Database}\\
\hline\hline 
SN & Type & Number of & Reference\\ 
Name && spectra &\\
\hline
SN~1981B & Ia       & 6  &\citet{Branch83}\\ 
SN~1986G & 91bg-like  & 5  &\citet{Phillips87}\\
SN~1990N & Ia       & 7  &\citet{Leibundgut91,Phillips92,Mazzali93}\\
SN~1991T & 91T-like & 9  &\citet{Filippenko92}\\
SN~1991bg & 91bg-like & 6  &\citet{Phillips92,Mazzali95}\\
SN~1992A & Ia       & 12 &\citet{Kirshner93}\\
SN~1994D & Ia       & 18 &\citet{Meikle96,Patat96}\\
SN~1996X & Ia       & 6  &\citet{Salvo01}\\  
SN~1997br & 91T-like & 10 &\citet{Li99}\\     
SN~1997cn & 91bg-like & 1  &\citet{turatto98}\\
SN~1998bu & Ia      & 12 &\citet{Jha99}\\    
SN~1999aa & 91T-like & 10 &\citet{Garavini04}\\
SN~1999by & 91bg-like & 14 &\citet{Garnavich04}\\
SN~1999ee & Ia      & 11 &\citet{Hamuy02}\\  
SN~2000cx & 91T-like & 24 &\citet{Li01c}\\    
SN~2002bo & Ia      & 12 &\citet{Benetti04}\\
SN~2003du & Ia      & 4  &\citet{Anupama05}\\
\hline
Total peculiar Ia = 79 & Total 'normal Ia' = 88 & Total = 167\\ 

\end{longtable}

\small
\begin{longtable}[c]{cccc}
\caption{\label{tab10} SN~Ib, Ic and II  template spectra in ${\cal SN}$-Fit Database}\\
\hline\hline
 SN & Type  & Number of & Reference\\ 
Name && spectra &\\
\hline
SN~1983N & Ib  & 2       & \citet{Harkness87}\\
SN~1984L & Ib & 5        & \citet{Harkness87}\\
SN~1987A & II\_pec & 4   & \citet{Pun95}\\    
SN~1987M & Ic & 1        & \citet{Filippenko90}\\
SN~1990B & Ic & 4        & \citet{Matheson01}\\
SN~1990U & Ic & 3       & \citet{Matheson01}\\
SN~1993J & IIb & 6       & \citet{Matheson00}\\
SN~1994I & Ic & 7        & \citet{Filippenko95}\\
SN~1997ef&  Ibc\_pec & 3 & \citet{Matheson01}\\
SN~1998bw& Ic & 7        & \citet{Patat01}\\  
SN~1998dt& Ib & 2        & \citet{Matheson01}\\
SN~1999di& Ib & 2        & \citet{Matheson01}\\
SN~1999dn& Ib & 3        & \citet{Matheson01}\\
SN~1999em& IIp & 11      & \citet{Hamuy01}\\  
SN~1999ex& Ibc & 3       &\citet{Hamuy02}\\  
SN~2000h & Ib & 1        & \citet{Branch02}\\ 
\hline
Total Ib/Ic = 43 & Total II =21 & Total Ib/Ic/II = 64\\
\end{longtable}

\clearpage

\clearpage

\end{document}